\documentclass[11pt]{article}

\usepackage[utf8]{inputenc}
\usepackage[iso-8859-1]{inputenx}
\usepackage{graphicx}
\usepackage[english]{babel}
\usepackage{amsmath,mathtools}
\usepackage{amssymb}
\usepackage{amsfonts}
\usepackage[margin=1in]{geometry}
\usepackage{hyperref}
\usepackage{float}
\usepackage{multirow}
\usepackage{latexsym}
\usepackage{caption}
\usepackage{babelbib}
\usepackage{latexsym}
\usepackage[T1]{fontenc}
\usepackage{tabularx}
\usepackage{dcolumn}
\usepackage{eurosym}
\usepackage{setspace}
\usepackage{color}
\usepackage[usenames,dvipsnames]{xcolor}
\usepackage{geometry}
\usepackage{multicol}
\usepackage{caption}
\usepackage{subfigure}
\usepackage{amsthm}
\usepackage{stmaryrd}
\usepackage{enumitem}
\usepackage{physics}
\usepackage{bbm}
\usepackage{bbding}
\usepackage{wasysym}
\usepackage{amsmath,accents}
\usepackage[dvipsnames]{xcolor}

\usepackage{mathrsfs}

\usepackage[T1]{fontenc}
\usepackage{tgbonum}
\usepackage{lmodern}

\usepackage{amssymb}
\usepackage{amssymb}%
\usepackage{pifont}%
\usepackage{float}

\restylefloat{table}

\numberwithin{equation}{section}

\newtheorem*{theorem*}{Theorem}

\let\OLDthebibliography\thebibliography
\renewcommand\thebibliography[1]{
  \OLDthebibliography{#1}
  \setlength{\parskip}{1.5pt}
  \setlength{\itemsep}{2pt plus 2ex}
}

\title{\textsf{\textbf{Asymptotic Symmetries of the Holst Action at Spatial Infinity: Including Supertranslations}}}
\author{
{\textrm{Sepideh Bakhoda}${}^{1,2}$\thanks{s.bakhoda@gmail.com}\;, \textrm{Hongguang Liu}${}^{2,3}$\thanks{liuhongguang@westlake.edu.cn}}
\\
\\
\textit{${}^1$Institute for Theoretical Physics and Cosmology,}\\
\textit{Zhejiang University of Technology, Hangzhou, 310023, China } \\
\textit{
${}^2$Institute for Theoretical Sciences and Department of Physics,}\\ 
\textit{Westlake University, Hangzhou 310030, China}\\
\textit{${}^3$Institute of Natural Sciences, Westlake Institute for Advanced Study, Hangzhou 310024, China}
}

\begin{document}

\maketitle
\sf{
\begin{abstract}
\rm{
We investigate the asymptotic symmetries of General Relativity at spatial infinity within the first-order formalism described by the Holst action. Employing the covariant phase space method, we propose a set of relaxed boundary conditions for the co-tetrad and Lorentz connection that admit the full Bondi-Metzner-Sachs (BMS) group, including non-trivial supertranslations, which are typically eliminated in standard treatments. We demonstrate that the logarithmic divergences appearing in the symplectic structure can be removed by imposing specific, symmetry-preserving parity conditions on the asymptotic fields without suppressing the supertranslation sector. A detailed analysis of the conserved charges reveals that the Holst term contributes non-trivially to the charge variations due to the linear growth of Lorentz generators. We show that the naive surface integrals for the Holst charges exhibit linear divergences arising from the rotation of the background tetrad. These divergences are successfully regularized by supplementing the asymptotic symmetry generator with a compensating internal Lorentz gauge transformation defined to preserve the background structure. The resulting charges are manifestly finite and integrable. Crucially, we prove that while the Holst modification shifts the charges associated with Lorentz boosts and rotations, it leaves the supertranslation charges identically invariant. This framework provides a consistent derivation of the full BMS algebra at spatial infinity in terms of Ashtekar-Barbero variables, offering new insights into the role of the Immirzi parameter in classical and quantum gravity.
}
\end{abstract}
}

\tableofcontents

\section{\textsf{Introduction}}
\rm{
A deep understanding of the fundamental symmetries of classical General Relativity is essential for elucidating the structure of spacetime and gravitational interactions, particularly in the physically significant context of asymptotically flat spacetimes \cite{Ashtekar:1978zani, Wald:1999wa, Geroch:1972up}. These symmetries underpin the definition of conserved quantities that characterize gravitational systems \cite{Regge:1974zd, Bondi:1962px, Sachs:1962wk} and reveal intricate structures at the boundaries of spacetime \cite{Strominger:2017zoo}. Such understanding is also pivotal for informing and guiding approaches to quantum gravity, where classical symmetries and their associated charges are expected to play fundamental roles in defining the quantum theory's states and observables \cite{Rovelli:2004tv, Thiemann:2007zz, Ashtekar:1981sf, Henneaux:1992ig}. The Ashtekar-Barbero variables \cite{Ashtekar:1986yd, BarberoG:1994eia, Immirzi:1996di}, central to Loop Quantum Gravity (LQG) \cite{Rovelli:2004tv, Thiemann:2007zz}, provide a unique framework for investigating these aspects. Concurrently, the study of asymptotic symmetries has revealed that the symmetry group of asymptotically flat spacetimes is not merely the Poincar\' e group, but the infinite-dimensional Bondi-Metzner-Sachs (BMS) group \cite{Bondi:1962px, Sachs:1962wk, Sachs:1962zzb}. This group, which incorporates supertranslations and potentially other enhancements like superrotations \cite{Barnich:2009se, Barnich:2010eb}, has profound implications. Understanding how the BMS group, initially discovered at null infinity, manifests at spatial infinity \cite{Ashtekar:1978zani, Henneaux:2018cst} is critical for a complete picture of gravitational dynamics, its conservation laws \cite{Strominger:2017zoo, He:2014laa}, and its broader implications for gravitational theory.

In a previous investigation \cite{Bakhoda:2023onetwo}, we explored novel boundary conditions for the Ashtekar-Barbero variables within the Hamiltonian formalism, specifically aiming to accommodate non-trivial supertranslations at spatial infinity. Standard parity conditions, such as those discussed by Regge and Teitelboim \cite{Regge:1974zd} or their analogues in the Ashtekar-Barbero framework \cite{Thiemann:1994NUE, Campiglia:2014eta}, typically ensure the finiteness of Poincar\' e charges but often lead to vanishing supertranslation charges. Our work in \cite{Bakhoda:2023onetwo} proposed alternative parity conditions that successfully yielded finite, integrable, and non-trivial supertranslation charges, while also ensuring a well-defined symplectic structure and the preservation of boundary conditions under hypersurface deformations, crucial consistency requirements for a Hamiltonian framework. However, a significant challenge remained: while the supertranslation charges were well-defined, the charges associated with asymptotic boosts and rotations were found to diverge under these new conditions. This outcome prevented the realization of the full BMS algebra at spatial infinity. This contrasted with the results obtained by Henneaux and Troessaert using ADM variables \cite{Henneaux:2018cst}, who successfully recovered the full BMS group by imposing specific off-shell conditions on the leading terms of the constraints. This discrepancy highlighted potential subtleties in the canonical equivalence of ADM and Ashtekar-Barbero variables at the boundary under such relaxed parity conditions, particularly concerning the surface terms critical for charge definitions \cite{Bakhoda:2023onetwo}.

The divergence of boost and rotation charges in our prior Hamiltonian analysis, despite the successful incorporation of supertranslations, motivates the present work to re-examine this problem from a covariant (Lagrangian) perspective. The covariant phase space formalism \cite{Crnkovic:1987pza, Lee:1990gr, Ashtekar:1990gc} offers an alternative route to defining asymptotic symmetries and their associated charges, which may naturally circumvent some of the difficulties encountered in the Hamiltonian approach or suggest different structural requirements for the boundary conditions. This paper will specifically investigate the asymptotics of the Holst action \cite{Holst:1995pc}, a first-order formulation of General Relativity, with the goal of identifying boundary conditions for the 4-dimensional co-tetrad $e_\mu^I$ and the Lorentz connection $\omega_\mu^{IJ}$ that accommodate the full BMS group, including supertranslations, at spatial infinity.
In this work, we restrict our attention to the strictly well-defined BMS$_4$ algebra, keeping the local conformal structure of the 2-sphere rigid. The potential enhancement to extended BMS algebras with singular superrotations is left for future exploration.

Addressing this problem using the \textbf{Hamiltonian approach}, as undertaken in \cite{Bakhoda:2023onetwo} and other works like \cite{Regge:1974zd, Henneaux:2018cst, Thiemann:1994NUE, Campiglia:2014eta}, generally involves these key steps:
\begin{itemize}
    \item Defining the canonical phase space variables (e.g., the 3-dimensional Ashtekar-Barbero connection $A_a^i$ and densitized triad $E_i^a$) on a spatial hypersurface $\Sigma$.
    \item Imposing fall-off conditions for these variables as spatial infinity ($r \to \infty$) on $\Sigma$ is approached.
    \item Defining parity conditions for the leading asymptotic parts of the canonical fields to ensure finiteness of the symplectic structure and charges.
    \item Ensuring the symplectic structure is well-defined and finite.
    \item Verifying that the chosen boundary conditions are preserved under hypersurface deformations (i.e., consistent with time evolution as generated by the Hamiltonian).
    \item Ensuring that the Hamiltonian generators of asymptotic symmetries (obtained from the constraints of the theory) are well-defined, finite, functionally differentiable, and integrable. This often involves careful treatment of surface terms and may involve imposing conditions on the asymptotic behavior of Lagrange multipliers or even on the leading terms of the constraints themselves \cite{Henneaux:2018cst}.
\end{itemize}

In contrast, the \textbf{covariant (Lagrangian) approach} that will be adopted in this work typically involves:
\begin{itemize}
    \item Starting with a 4-dimensional action principle, such as the Holst action for General Relativity.
    \item Defining the configuration space in terms of 4-dimensional fields (e.g., the co-tetrad $e_\mu^I$ and the Lorentz connection $\omega_\mu^{IJ}$ as independent fields).
    \item Specifying the conditions for asymptotic flatness at spatial infinity ($i^0$), which translates to precise fall-off rates for $e_\mu^I$ towards a background Minkowski tetrad and for $\omega_\mu^{IJ}$ towards the corresponding flat connection.
    \item Ensuring the action is well-defined under these conditions. First-order actions like the Palatini or Holst action are often advantageous as they may not require infinite counter-terms or conventional Gibbons-Hawking-York type boundary terms for a well-posed variational principle if field fall-offs are appropriately chosen \cite{Ashtekar:2008jw}.
    \item Deriving the pre-symplectic current from the Lagrangian and constructing the symplectic structure on the covariant phase space (the space of solutions to the field equations satisfying the boundary conditions). The finiteness of this structure and the absence of flux leakage at spatial infinity are paramount.
    \item Identifying asymptotic symmetry vector fields $\xi^\mu$ as those preserving the asymptotic form of the fields and the boundary conditions. These should include supertranslations to aim for the BMS group.
    \item Constructing the charges associated with these symmetries. This is typically achieved through surface integrals at spatial infinity, derived from the symplectic structure and the action, often utilizing Noether's second theorem or related covariant phase space techniques \cite{Lee:1990gr, Wald:1999wa, Iyer:1994ys}.
\end{itemize}
This paper will focus on the latter approach, seeking to establish a consistent set of boundary conditions for the tetrad and Lorentz connection within the first-order covariant framework that successfully incorporates the full BMS group, including supertranslations, at spatial infinity.
\\
\\
This work is organized as follows:
\\
In Section \ref{Section 2}, we review the first-order Holst action, define its covariant phase space, and derive the corresponding pre-symplectic current. 
\\
Section \ref{Sec.3} details the asymptotic structure of spacetime at spatial infinity, specifying the radial-hyperbolic coordinate system and the asymptotic expansions for the co-tetrad and Lorentz connection. 
\\
In Section \ref{Section 4}, we evaluate the pre-symplectic structure and explicitly isolate the logarithmic divergences that arise when independent supertranslations are accommodated. 
\\
To resolve this, Section \ref{Section5} introduces a consistent set of parity conditions on the asymptotic boundary data that regularize the symplectic form while being perfectly preserved under the full BMS group. 
\\
Section \ref{sec:charges} is dedicated to the explicit construction of the conserved charges. We separately analyze the Palatini and Holst contributions, demonstrating the necessity of an internal Lorentz gauge compensator to extract finite Holst charges, and proving the invariance of supertranslation charges under the Holst modification. 
\\
In Section \ref{Section 7}, we rigorously establish the closure of the extended asymptotic BMS$_4$ algebra using the modified Barnich-Troessaert bracket, proving the consistency of the field-dependent translations and internal gauge compensators. 
\\
Section \ref{sec:discussion} provides a discussion on the broader physical implications of our framework, including its relationship to edge modes, holographic duality, and the self-dual limit. 
\\
Finally, Section \ref{sec:conclusion} summarizes our findings and concludes the paper. 
\\
Detailed derivations of the field transformations under the asymptotic symmetries are relegated to Appendix \ref{app:field_transformations}.
}

\section{\textsf{The Holst Action and its Covariant Phase Space}}\label{Section 2}

This section is dedicated to reviewing the necessary background formalism. We begin by introducing the Holst action and its properties within the first-order framework. We then define its corresponding covariant phase space and derive the pre-symplectic current, which is the foundational element for defining conserved charges. The notation used throughout this paper will also be established here.

\subsection{\textsf{Holst Action}}
We consider the Holst action for General Relativity on a 4-dimensional spacetime manifold $\mathcal{M}$. The fundamental variables are the co-tetrad (or frame field) 1-form $e^I = e_\mu^I dx^\mu$ and the $so(1,3)$ Lorentz connection 1-form $\omega^{IJ} = \omega_\mu^{IJ} dx^\mu$. Here, $I, J, K, L$ run from $0$ to $3$ are internal Lorentz indices, raised and lowered with the Minkowski metric $\eta_{IJ} = \text{diag}(-1, 1, 1, 1)$ (or an appropriate choice for the signature). Spacetime indices $\mu, \nu, \rho, \sigma \in \{t,x,y,z\}$. The spacetime metric is reconstructed from the co-tetrad via $g_{\mu\nu} = e_\mu^I e_\nu^J \eta_{IJ}$. The connection $\omega_\mu^{IJ}$ is antisymmetric in its internal indices, $\omega_\mu^{IJ} = -\omega_\mu^{JI}$.
The curvature 2-form of the connection $\omega^{IJ}$ is $F^{IJ} = d\omega^{IJ} + \omega^I{}_K \wedge \omega^{KJ}$.
The Holst action is given by \cite{Holst:1995pc, Corichi:2010aaa}:
\begin{equation}
S[e,\omega] = -\frac{1}{2\kappa} \int_{\mathcal{M}} \left( \Sigma_{IJ} \wedge F^{IJ} + \frac{1}{\beta} \Sigma_{IJ} \wedge \star F^{IJ} \right)
\label{eq:HolstActionForms}
\end{equation}
where $\kappa = 8\pi G$, $\beta$ is the Barbero-Immirzi parameter, and $\Sigma_{IJ} = \frac{1}{2}\epsilon_{IJKL} e^K \wedge e^L$. The term $\star F^{IJ}$ denotes the Hodge dual of $F^{IJ}$ with respect to the internal indices, i.e., $\star F^{IJ} = \frac{1}{2}\epsilon^{IJ}{}_{KL}F^{KL}$. The first term in the action \eqref{eq:HolstActionForms} is the standard Palatini action. The second term, which is proportional to $1/\beta$, is known as the Holst term. A key feature of the Holst term is that it does not alter the classical equations of motion, which remain equivalent to the vacuum Einstein equations. On solutions, the connection $\omega_\mu^{IJ}$ becomes the unique torsion-free, metric-compatible connection \cite{Holst:1995pc}.

When the field equations derived from the variation of $S[e,\omega]$ with respect to $\omega_\mu^{IJ}$ are imposed, they imply that the torsion 2-form $T^I = de^I + \omega^I{}_J \wedge e^J$ vanishes. The condition $T^I=0$ means that $\omega_\mu^{IJ}$ are no longer independent fields but are determined by the co-tetrad $e_\mu^I$ and its first derivatives. These are the Ricci rotation coefficients. An explicit formula for $\omega_\mu{}^{IJ}$ satisfying the torsion-free condition is \cite{Rovelli:2004tv, Misner:1973prb}:
\begin{equation}
\omega_\mu{}^{IJ}(e) = \frac{1}{2} e^{\nu I} (\partial_\mu e_{\nu}{}^J - \partial_\nu e_{\mu}{}^J) - \frac{1}{2} e^{\nu J} (\partial_\mu e_{\nu}{}^I - \partial_\nu e_{\mu}{}^I) - \frac{1}{2} e_\mu^K e^{\nu I} e^{\lambda J} (\partial_\nu e_{\lambda K} - \partial_\lambda e_{\nu K})
\label{eq:RicciRotationCoeffs}
\end{equation}
Here, $e^\nu_I$ is the inverse tetrad ($e^\nu_I e_\mu^I = \delta^\nu_\mu$, $e^\nu_I e_\nu^J = \delta_I^J$), $e_{\nu K} = e_\nu^L \eta_{LK}$, and internal indices on $e$ are raised with $\eta^{IJ}$.

The covariant phase space $\Gamma$ for this theory consists of smooth field configurations $(e_\mu^I, \omega_\mu^{IJ})$ on the 4-manifold $\mathcal{M}$ (assumed diffeomorphic to $\mathbb{R}^4$) that are solutions to the full field equations (implying Einstein's equations and the torsion-free condition; thus $\omega$ is given by Eq. \eqref{eq:RicciRotationCoeffs} in terms of $e$). We will be particularly interested in the subspace of $\Gamma$ corresponding to solutions that are asymptotically flat at spatial infinity and admit supertranslations.

\subsection{\textsf{Pre-symplectic Potential and Current}}
 So far, we have considered the bulk part of the action. For a manifold $\mathcal{M}$ with a boundary $\partial\mathcal{M}$, the variation of the bulk action $S_{\text{bulk}}$ produces surface terms due to the presence of derivatives. Taking the variation of Eq. \eqref{eq:HolstActionForms} and integrating by parts yields
\[ \delta S_{\text{bulk}} = -\frac{1}{2\kappa}\int_{\mathcal{M}} (\text{EOM terms}) \cdot \delta\phi - \frac{1}{2\kappa} \int_{\partial\mathcal{M}} \Sigma_{IJ} \wedge \left( \delta\omega^{IJ} + \frac{1}{\beta} \delta(\star\omega^{IJ}) \right) \]
Here, $\phi$ collectively denotes the fundamental fields of the theory, i.e. $\phi=(e_\mu^I, \omega_\mu^{IJ})$.
The boundary term contains variations of the connection, $\delta\omega^{IJ}$, which means the action principle is not well-posed if we only fix the co-tetrad $e^I$ on the boundary. To remedy this, a suitable boundary term must be added to the action and the total action becomes \cite{Corichi:2010aaa}
\begin{equation}
S_{\text{total}}[e,\omega] = -\frac{1}{2\kappa} \int_{\mathcal{M}}  \Sigma_{IJ} \wedge \left( F^{IJ} + \frac{1}{\beta} \star F^{IJ} \right) + \frac{1}{2\kappa} \int_{\partial\mathcal{M}} \Sigma_{IJ}\wedge\left(\omega^{IJ}+\frac{1}{\beta}\star \omega^{IJ}\right)
\label{eq:HolstActionTotal_Sec2}
\end{equation}
The variation of this total action is:
\[ \delta S_{\text{total}} = -\frac{1}{2\kappa} \int_{\mathcal{M}} (\text{EOM terms}) \cdot \delta\phi + \frac{1}{2\kappa}\int_{\partial\mathcal{M}}\delta\Sigma_{IJ}\wedge\left(\omega^{IJ}+\frac{1}{\beta}\star \omega^{IJ}\right) \]
The bulk term vanishes on-shell, and the remaining boundary term defines the pre-symplectic potential 1-form (on field space) 
\begin{equation}
\Theta(\delta) = \frac{1}{2\kappa} \delta\Sigma_{IJ}\wedge\left(\omega^{IJ}+\frac{1}{\beta}\star\omega^{IJ}\right)
\label{eq:SymplecticPotentialCWE_main_sec2}
\end{equation}
The symplectic current 3-form $J(\delta_1, \delta_2) = \delta_1 \Theta(\delta_2) - \delta_2 \Theta(\delta_1)$ is then given by 
\begin{equation}
J(\delta_{1},\delta_{2})=-\frac{1}{2\kappa}\left[\delta_{1}\Sigma_{IJ}\wedge\delta_{2}(\omega^{IJ}+\frac{1}{\beta}\star\omega^{IJ})-\delta_{2}\Sigma_{IJ}\wedge\delta_{1}(\omega^{IJ}+\frac{1}{\beta}\star \omega^{IJ})\right]
\label{eq:SymplecticCurrentKinematic_main_sec2}
\end{equation}
This is the full off-shell symplectic current. When evaluated on solutions where the torsion-free condition holds (half on-shell), the contribution from the Holst term can be written as a total divergence, and the current simplifies to \cite{Corichi:2010aaa}
\begin{align}
J(\delta_{1},\delta_{2})&=-\frac{1}{2\kappa}\left[\delta_{1}\Sigma_{IJ}\wedge\delta_{2}\omega^{IJ}-\delta_{2}\Sigma_{IJ}\wedge\delta_{1}\omega^{IJ}\right]+\frac{1}{\kappa\beta}d(\delta_{1}e^{I}\wedge\delta_{2}e_{I})\nonumber \\ 
& =:
J_P(\delta_{1},\delta_{2}) + J_H(\delta_{1},\delta_{2})
\label{eq:SymplecticCurrentHalfShell_main_sec2}
\end{align}
The first term is the Palatini symplectic current $J_P$, and the second is the exact form arising from the Holst modification $J_H$. 
The pre-symplectic structure is defined by integrating this current over a Cauchy surface, 
\begin{equation}
    \Omega(\delta_{1},\delta_{2})=\int_{\Sigma}J(\delta_{1},\delta_{2})
\label{eq:Symplectic_Form_General}    
\end{equation} 
A fundamental requirement for this structure to be physically meaningful is that it must be independent of the choice of Cauchy surface $\Sigma$. By Stokes' theorem, this is equivalent to demanding that there is no ``leakage'' of the symplectic flux at spatial infinity. As noted in \cite{Ashtekar:2008jw}, the issue of whether the boundary conditions ensure the flux across the timelike cylinder at infinity, $\int_{\tau_\infty}J$, vanishes is delicate and crucial. If this flux does not vanish, we would not obtain a well-defined symplectic structure on the phase space $\Gamma$.

Therefore, a primary task ahead is to analyze the asymptotic behavior of the symplectic current \eqref{eq:SymplecticCurrentHalfShell_main_sec2} using the field expansions from Section \ref{Sec.3}. We must then identify boundary conditions (specifically, parity conditions on the asymptotic data) that guarantee the vanishing of this flux. This will ensure that our phase space is equipped with a consistent and conserved symplectic structure, providing a solid foundation for defining the charges associated with the BMS group.

\section{\textsf{Asymptotic Structure}}\label{Sec.3}

To study supertranslations at spatial infinity, we need to define a suitable class of asymptotically flat spacetimes and the behavior of our fundamental fields $e_\mu^I$ and $\omega_\mu^{IJ}$ in this asymptotic region. Unlike approaches that impose strict conditions to isolate the Poincar\' e group \cite{Ashtekar:2008jw, Corichi:2010aaa}, our goal is to allow for a broader class of asymptotic behaviors consistent with the BMS group.

\subsection{\textsf{Asymptotic Flatness and Coordinates}}\label{Section3.1}
We consider spacetimes that are asymptotically flat at spatial infinity ($i^0$). We use Cartesian coordinates $x^\mu$ in the bulk, associated with a background Minkowski metric $\eta_{\mu\nu}$. For analyzing the asymptotic region, we employ a \textit{radial-hyperbolic} coordinate system $(\rho, \Phi^A)$, where $r$ is a radial coordinate defined by $\rho^2 = \eta_{\mu\nu}x^\mu x^\nu$ (assuming $\rho^2>0$), which tends to infinity at spatial infinity ($\rho \to \infty$). The coordinates $\Phi^A=(\chi, \theta, \phi)$ (with $A=\chi, \theta, \phi$) are angles that parameterize the asymptotic 3-dimensional hyperboloids $\tau_\rho$ of constant $\rho$. The coordinate $\chi$ here is a hyperbolic ``time angle''.

The general metric form of an asymptotically flat space-time is
\begin{equation}
ds^2 = \left(1+\frac{2\sigma(\Phi)}{\rho}\right)d\rho^2 + \rho^2 \left(h_{AB}(\Phi) + \frac{{}^1h_{AB}(\Phi)}{\rho}\right)d\Phi^A d\Phi^B + o(\rho^{-1})
\label{eq:GeneralMetricHyperbolic_r_revised}
\end{equation}
Here, $\sigma(\Phi)$ and ${}^1h_{AB}(\Phi)$ are functions of angles $\Phi=(\chi,\theta, \phi)$ only. The metric $h_{AB}(\Phi)$ is the metric on the unit 3-hyperboloid (e.g., for a unit timelike hyperboloid in Minkowski space, $h_{AB}d\Phi^A d\Phi^B = -d\chi^2 + \cosh^2\chi(d\theta^2 + \sin^2\theta d\phi^2)$). The metric \eqref{eq:GeneralMetricHyperbolic_r_revised} has vanishing $g_{\rho A}$ components at the leading orders shown\footnote{It is instructive to relate these radial-hyperbolic coordinates to standard formulations. The limit $\rho \to \infty$ approaches spatial infinity along slices of constant hyperbolic angle $\chi$. For instance, on a constant-time slice $t = \rho\sinh\chi$, this differs fundamentally from the approach to null infinity, where the retarded time $u = t - r$ is held constant. Consequently, this radial parameterization naturally bridges the ADM Cauchy formulation in the bulk with the rigid boundary frame at spatial infinity.}.

Before proceeding, it is important to clarify the relationship between the boundary conditions we intend to use and those employed in  \cite{Ashtekar:2008jw, Corichi:2010aaa}. To obtain a unique Poincar\' e group at spatial infinity, those studies imposed two key restrictions on the asymptotic data:
\begin{enumerate}
    \item A parity condition on the mass aspect, requiring $\sigma(\Phi)$ to be an \textit{even} function under the antipodal map on the unit hyperboloid (a map that relates a point on the hyperboloid to the point directly opposite it). This condition is designed to eliminate the freedom to perform so-called logarithmic translations.
    \item A second condition relating the angular metric perturbation to the mass aspect: ${}^1h_{AB}(\Phi) = -2\sigma(\Phi)h_{AB}(\Phi)$. This stronger condition effectively removes the freedom of supertranslations, thereby restricting the asymptotic symmetry group to the Poincar\' e group.
\end{enumerate}
In this paper, our goal is to investigate the full BMS group, including supertranslations. Therefore, while we will retain the first condition that $\sigma(\Phi)$ is even to prevent logarithmic ambiguities, we will \textit{not} impose the second condition. Instead, we will treat ${}^1h_{AB}(\Phi)$ as a general, independent field parameterizing the asymptotic data. Our objective is to determine what new boundary conditions (e.g., parity conditions on the components of ${}^1h_{AB}$) are required to ensure that the symplectic structure is well-defined (i.e., has no flux leakage at infinity) and that the charges for all generators of the BMS group are finite and integrable.

To facilitate our calculations, we now define some key geometric quantities that will be used throughout the paper. These are constructed from the background Cartesian coordinates $x^\mu$:
\begin{itemize}
    \item \textbf{Unit radial vector} (Cartesian components): $\hat{x}_\mu(\Phi) = \eta_{\mu\nu}x^\nu/\rho$. Its internal counterpart is $\hat{x}^I(\Phi)$.
    \item \textbf{Basis covectors for the angular directions} on the unit hyperboloid (Cartesian components): $\hat{e}^A_\mu(\Phi)$, for $A \in \{\chi, \theta, \phi\}$. These are tangent to the unit hyperboloid and orthogonal to $\hat{x}_\mu$. Their internal counterparts (as vectors) are $\hat{e}^{A I}(\Phi)$.
\end{itemize}

\subsection{\textsf{Asymptotic Coordinate Transformations}}\label{sec3.2}
The symmetries of an asymptotically flat spacetime are generated by vector fields $\xi^\mu$ whose action preserves the asymptotic structure laid out in Section \ref{Section3.1}. We are interested in the vector fields corresponding to the BMS group, which, in the asymptotic limit, take the following form in Cartesian coordinates:
\begin{itemize}
    \item \textbf{Lorentz Transformations}: $\xi^\mu = M^\mu{}_\nu x^\nu$, where $M_{\mu\nu} = -M_{\nu\mu}$ is a constant antisymmetric tensor. This vector field, which grows linearly with the radial coordinate, generates the rotations (for spatial $M_{ab}$) and boosts (for time-space $M_{ta}$) of the Poincar\'e group.

    \item \textbf{Supertranslations}: $\xi^\mu = S^\mu(\Phi)$. This is the key feature of the BMS group, extending standard translations to include angle-dependent shifts. The functions $S^\mu(\Phi)$ depend only on the asymptotic angles $\Phi$ and are of order $\mathcal{O}(1)$ in $\rho$. The zero modes (or $\ell=0,1$ spherical harmonic modes) of $S^\mu(\Phi)$ correspond to the ordinary Poincar\' e translations.
\end{itemize}
For the subsequent analysis of the symplectic structure and the associated charges, it is instructive to decompose the general asymptotic Poinca\'e generator $\xi^\mu$ into its components normal and tangential to the asymptotic hyperboloids. This is achieved by projecting $\xi^\mu$ onto the orthonormal basis $\{\hat{x}^\mu, \hat{e}^\mu_A\}$. The general decomposition takes the form:
\begin{equation}
    \xi^\mu = C_\rho(\Phi) \hat{x}^\mu + C^A(\Phi) \hat{e}^\mu_A
\end{equation}
The radial component, $C_\rho$, is given by the projection of the translation vector onto the radial direction, as the Lorentz part vanishes due to the anti-symmetry of $M^\mu_\nu$:
\begin{equation}
    C_\rho = \xi^\mu \hat{x}_\mu = S^\mu \hat{x}_\mu =:S
\end{equation}
The tangential components, $C^A$, contain contributions from both the Lorentz and translation parts of the generator:
\begin{equation}
    C^A = \xi^\mu \hat{e}^A_\mu = (M^\mu_\nu x^\nu)\hat{e}^A_\mu + S^\mu \hat{e}^A_\mu
\end{equation}
The Lorentz contribution can be further decomposed into parts generated by boosts and spatial rotations. We define the boost generator $B^A(\Phi)$ and the rotation generator $R^A(\Phi)$ on the hyperboloid as the projections of the corresponding constant Lorentz matrix $M^\mu_\nu$. These are given by 
\begin{align}
    B^A &:= M^t_a (\hat{x}^a \hat{e}^A_t - \hat{x}^t \hat{e}^{A a}) \\
    R^A &:= M^b_a \hat{x}^a \hat{e}^A_b
\end{align}
where the indices $a, b$ run over spatial Cartesian coordinates.
The vector fields $B^A$ and $R^A$ are Killing vectors of the hyperbolic metric $h_{AB}$, i.e.
\begin{equation}
    \mathcal{L}_{B} h_{AB}=0, \quad  \mathcal{L}_{R} h_{AB}=0
\end{equation}
It is worth noting that a direct consequence of this definition and the geometric properties of the basis vectors is that the $\chi$-component of the rotation generator vanishes identically, i.e., $R^\chi = 0$. Using these definitions, the tangential components can be written as:
\begin{equation}\label{definition of X}
    C^A = \rho X^A + S^A
\end{equation}
where $S^A:= S^\mu \hat{e}^A_\mu$ and $X^A:= B^A+R^A$.

This decomposition provides a clear separation of the different physical transformations within the geometric framework of the asymptotic hyperboloid.
Our primary objective in this work is to establish a framework where the charges associated with this full set of transformations, including the non-trivial supertranslations, are well-defined and finite.

\subsection{\textsf{Asymptotic Behavior of Tetrad and Connection}}

For the metric to be asymptotically flat as defined in Eq. \eqref{eq:GeneralMetricHyperbolic_r_revised}, the co-tetrad $e_\mu^I$ must approach the flat Minkowski space tetrad, ${}^o e_\mu^I = \delta_\mu^I$, as $\rho \to \infty$. The condition that the metric deviation $g_{\mu\nu} - \eta_{\mu\nu}$ falls off as $\mathcal{O}(1/\rho)$ implies that the co-tetrad itself, $e_\mu^I$, should admit an asymptotic expansion starting with a $1/\rho$ correction. Therefore, we expand the Cartesian components of the co-tetrad as
\begin{equation}
e_\mu^I(x) = \delta_\mu^I + \frac{1}{\rho}{}^1e_\mu^I(\Phi) + \frac{1}{\rho^2}{}^2e_\mu^I(\Phi) + o(\rho^{-2})
\label{eq:tetrad_exp_cartesian_r_revised}
\end{equation}
where the coefficients ${}^ne_\mu^I(\Phi)$ depend only on the angles $\Phi = \Phi(x^\mu/\rho)$.

Note that in our framework, we consider the asymptotic fields $\sigma(\Phi)$ and ${}^1h_{AB}(\Phi)$ from the metric expansion \eqref{eq:GeneralMetricHyperbolic_r_revised} to be the fundamental physical quantities on which our parity conditions will be imposed. It is for this reason that, in the remainder of this section, we must derive the leading terms in the asymptotic expansions of the co-tetrad and the connection in terms of these quantities.

To find ${}^1e_\mu^I$ in terms of $\sigma(\Phi)$ and ${}^1h_{AB}(\Phi)$,  we first need to transform the Cartesian tetrad components \eqref{eq:tetrad_exp_cartesian_r_revised} into radial-hyperbolic coordinates and then substitute them into the relation $g_{\mu\nu} = e_\mu^I e_\nu^J \eta_{IJ}$ to solve for ${}^1e_\mu^I(\Phi)$. This procedure involves:

\begin{enumerate}
    \item Transforming the Cartesian tetrad components $e_\mu^{\text{Cart},I}$ to radial-hyperbolic components $e_\lambda^{\text{hyp},I} = \frac{\partial x^\mu}{\partial y^\lambda} e_\mu^{\text{Cart},I}$. Let the leading correction to $e_\rho^I$ be $\frac{1}{\rho}{}^{1}e_\rho^I(\Phi)$ and the leading ($\mathcal{O}(1)$) correction to $e_{A}^I$ be ${}^{1}e_{A}^I(\Phi)$. These are given by
    \begin{align*}
        {}^{1}e_\rho^I(\Phi) &= \hat{x}^\mu(\Phi) {}^1e_\mu^{\text{Cart},I}(\Phi) \\
        {}^{1}e_{A}^I(\Phi) &= \hat{e}_A^\mu(\Phi) {}^1e_\mu^{\text{Cart},I}(\Phi)
    \end{align*}
    where $\hat{x}^\mu$ is the Cartesian unit radial vector and $\hat{e}_A^\mu$ are Cartesian components of unit angular basis vectors.
    \item Using $g_{\mu\nu} = e_\mu^I e_\nu^J \eta_{IJ}$ we get the following conditions:
    \begin{itemize}
        \item $g_{\rho \rho} = 1 + \frac{2}{\rho} (\hat{x}_I {}^{1}e_\rho^I) + \mathcal{O}(\rho^{-2})$. Matching with $1 + \frac{2\sigma}{\rho}$ gives
        \begin{equation} \hat{x}_I (\hat{x}^\mu \; {}^1e_\mu^{\text{Cart},I}(\Phi)) = \sigma(\Phi) \label{eq:condA_r} \end{equation}
        \item The $\mathcal{O}(1)$ term of $g_{\rho A} = \hat{x}_I {}^{1}e_{A}^I + {}^{1}e_{\rho I} \hat{e}_{A}^I$. Since $g_{\rho A} = \mathcal{O}(\rho^0)$, this must vanish
        \begin{equation}
        \hat{x}_I (\hat{e}_{A}^\mu {}^1e_\mu^{\text{Cart},I}(\Phi)) + (\hat{x}^\nu {}^1e_\nu^{\text{Cart}}(\Phi))_I \hat{e}_{A}^I(\Phi) = 0 \label{eq:condB_r} \end{equation}
        \item The $\mathcal{O}(\rho)$ term of $g_{AB} = \rho^2 (\hat{e}_{A I} {}^{1}e_{B}^I + {}^{1}e_{A I} \hat{e}_B^I )$. Comparing with $\rho^2 \; {}^1h_{AB}(\Phi)$
        \begin{equation} 
        \hat{e}_{A I} (\hat{e}_{B}^\mu {}^1e_\mu^{\text{Cart},I}(\Phi)) + (\hat{e}_A^\nu {}^1e_\nu^{\text{Cart}}(\Phi))_I \hat{e}_{B}^I(\Phi) = {}^1h_{AB}(\Phi) \label{eq:condC_r} \end{equation}
    \end{itemize}

\item Decomposing ${}^1e_\mu^{\text{Cart},I}(\Phi) = A(\Phi) \hat{x}_\mu \hat{x}^I + B_A(\Phi) \hat{x}_\mu \hat{e}^{AI} + C^A(\Phi) \hat{e}_{A\mu} \hat{x}^I + D_{AB}(\Phi) \hat{e}^A_\mu \hat{e}^{B I}$ and solving for $A, B_A, C^A, D_{AB}$ using conditions \eqref{eq:condA_r}, \eqref{eq:condB_r}, \eqref{eq:condC_r}:
\begin{itemize}
    \item From \eqref{eq:condA_r}, we uniquely fix the radial component: $A(\Phi) = \sigma(\Phi)$.
    \item From \eqref{eq:condB_r}, we obtain the relation $C_A(\Phi) + B_A(\Phi) = 0$. The geometric information from the metric only fixes this combination. To completely determine the tetrad, we must fix the local Lorentz gauge freedom. We choose the symmetric (radial) gauge by setting the internal radial leg of the tetrad to have no angular components, which implies $B_A(\Phi) = 0$. Consequently, $C_A(\Phi) = 0$.
    \item From \eqref{eq:condC_r}, we get $D_{BA}(\Phi) + D_{AB}(\Phi) = {}^1h_{AB}(\Phi)$. By further exploiting the remaining local Lorentz frame choice (spatial rotations), we can choose the symmetric part, setting $D_{AB}(\Phi) = \frac{1}{2}{}^1h_{AB}(\Phi)$ and its antisymmetric part to zero.
\end{itemize}
\end{enumerate}

This yields the expression for the Cartesian coefficient ${}^1e_\mu^I(\Phi)$:
\begin{equation}
{}^1e_\mu^I(\Phi) = \sigma(\Phi) \hat{x}_\mu(\Phi) \hat{x}^I(\Phi) + H_\mu^I(\Phi)
\label{eq:1e_derived_r}
\end{equation}
where we define the purely tangential part (in both spacetime Cartesian and internal indices) as
\begin{equation}
H_\mu^I(\Phi) := \frac{1}{2} {}^1h_{AB}(\Phi) \hat{e}^A_\mu(\Phi) \hat{e}^{BI}(\Phi)
\label{eq:H_def_r}
\end{equation}

Now, we consider the asymptotic expansion of the Lorentz connection $\omega_\mu^{IJ}$. This connection is determined by the tetrad via the Ricci rotation coefficient formula \eqref{eq:RicciRotationCoeffs}. 
When these are inserted into the formula \eqref{eq:RicciRotationCoeffs} for $\omega_\mu^{IJ}$, each term will involve at least one derivative of a tetrad component\footnote{
In substituting the tetrad expansion \eqref{eq:tetrad_exp_cartesian_r_revised} into \eqref{eq:RicciRotationCoeffs}, one notes that the leading term of $e_\mu^I$ is $\delta_\mu^I$ (constant), so its derivatives $\partial_\nu \delta_\mu^I$ are zero and the first derivatives of the $1/\rho$ correction term $\frac{1}{\rho}{}^1e_\nu^J(\Phi)$ are
    $$ \partial_\lambda \left(\frac{1}{\rho}{}^1e_\nu^J(\Phi)\right) = \frac{\partial_\lambda ({}^1e_\nu^J(\Phi))}{\rho} - \frac{{}^1e_\nu^J(\Phi) (\partial_\lambda \rho)}{\rho^2} $$
    Since $\partial_\lambda ({}^1e_\nu^J(\Phi)) \sim \mathcal{O}(1/\rho)$ (as ${}^1e_\nu^J$ depends on $\Phi = x^\alpha/\rho$) and $\partial_\lambda \rho = \hat{x}_\lambda(\Phi) \sim \mathcal{O}(1)$, the entire expression $\partial_\lambda (\frac{1}{\rho}{}^1e_\nu^J(\Phi))$ is $\mathcal{O}(1/\rho^2)$.
}.
Since the derivatives of the background tetrad $\delta_\mu^I$ are zero, the leading contributions to $\omega_\mu^{IJ}$ will come from terms like $\delta^{\nu I} \partial_\mu (\frac{1}{\rho}{}^1e_\nu^J)$, which are $\mathcal{O}(1/\rho^2)$.
Therefore, the coefficients of $\rho^0$ and $1/\rho$ in the expansion of the Cartesian components $\omega_\mu^{IJ}$ must vanish
\begin{equation}
{}^o\omega^{IJ}_\mu(\Phi) = 0 \quad \text{and} \quad {}^1\omega^{IJ}_\mu(\Phi) = 0
\end{equation}
Thus, the Cartesian connection expansion starts as:
\begin{equation}
\omega_\mu^{IJ}(x) = \frac{1}{\rho^2}{}^2\omega_\mu^{IJ}(\Phi) + \frac{1}{\rho^3}{}^3\omega_\mu^{IJ}(\Phi) + o(\rho^{-3})
\label{eq:omega_exp_cartesian_r}
\end{equation}
The coefficient ${}^2\omega_\mu^{IJ}(\Phi)$ is obtained by collecting all $\mathcal{O}(1/\rho^2)$ terms from substituting \eqref{eq:tetrad_exp_cartesian_r_revised} into Eq. \eqref{eq:RicciRotationCoeffs}. This leads to the expression
\begin{equation}
{}^2\omega_\mu^{IJ}(\Phi) = 2 \left[ \mathcal{D}^{[J} ({}^1e_\mu^{I]}(\Phi)) - {}^1e_\mu^{[I}(\Phi)\hat{x}^{J]}(\Phi) \right]
\label{eq:2omega_general_r}
\end{equation}
where ${}^1e_\mu^I(\Phi)$ is given by Eq. \eqref{eq:1e_derived_r}, and $\mathcal{D}^J F(\Phi) = \rho (\eta^{JK}\partial_K F(\Phi(x/\rho)))$ is the $\mathcal{O}(1)$ angular derivative operator (with $\partial_K = \partial/\partial x^K$).
Substituting Eq. \eqref{eq:1e_derived_r} into Eq. \eqref{eq:2omega_general_r} and using $H_\mu^I(\Phi)$ from Eq. \eqref{eq:H_def_r}
\begin{equation}\label{eq:2omega_final_explicit_r}
{}^2\omega_\mu^{IJ}(\Phi) = 2\left[(\mathcal{D}^{[J}\sigma)\hat{x}_\mu\hat{x}^{I]} + \sigma\delta_\mu^{[J}\hat{x}^{I]}\right] + 2\mathcal{D}^{[J}H_\mu^{I]}(\Phi) - 2H_\mu^{[I}(\Phi)\hat{x}^{J]}(\Phi)
\end{equation}
This expression gives the leading non-vanishing coefficient of the connection in terms of the asymptotic metric data $\sigma(\Phi)$ and ${}^1h_{AB}(\Phi)$ (via $H_\mu^I$), consistent with allowing supertranslations.

\section{\textsf{Pre-Symplectic Structure and its Conservation}}\label{Section 4}
In this section, we try to construct the pre-symplectic form \eqref{eq:Symplectic_Form_General}, which is the central element for defining the covariant phase space of the theory. The structure of the current \eqref{eq:SymplecticCurrentHalfShell_main_sec2}, evaluated on the space of solutions to the field equations, is fundamental to demonstrating that the symplectic structure is well-defined.

We begin with the half-on-shell symplectic current 3-form for the Holst action, given in Eq. \eqref{eq:SymplecticCurrentHalfShell_main_sec2}.
A crucial property for defining a conserved pre-symplectic structure is that this current must be closed, i.e., its exterior derivative must vanish ($dJ=0$) on the space of solutions. The Holst part ($J_H$) is already written as an exact form. Because the exterior derivative is nilpotent, the Holst part of the current is closed identically: $dJ_H = 0$. The closure of the Palatini part ($J_P$) is a non-trivial result that holds on the space of solutions. One can verify that $J_P$ is closed by using the fact that the fields $(e, \omega)$ satisfy the equations of motion and the variations $(\delta e, \delta \omega)$ satisfy the linearized equations of motion \cite{Ashtekar:2008jw}. The proof involves taking the exterior derivative $dJ_P$ and showing that the resulting terms cancel out when the on-shell conditions and the Bianchi identities are applied.
Since both parts are closed on the space of solutions, the total symplectic current is closed $dJ=0$.
Now, following the procedure in \cite{Ashtekar:2008jw}, let us consider a 4-dimensional region of spacetime $\tilde{\mathcal{M}}$, which is bounded by compact portions of two Cauchy surfaces, $\tilde{\Sigma}_1$ and $\tilde{\Sigma}_2$, and a timelike 3-surface $\tau$ that connects their boundaries ($\partial\tilde{\Sigma}_1$ and $\partial\tilde{\Sigma}_2$). The complete boundary of this region is $\partial\tilde{\mathcal{M}} = \tilde{\Sigma}_2 - \tilde{\Sigma}_1 + \tau$. Therefore, using the Stokes' theorem we have
\begin{equation}
0= \int_{\tilde{\mathcal{M}}}dJ=\int_{\partial\tilde{\mathcal{M}}} J = \int_{\tilde{\Sigma}_2} J - \int_{\tilde{\Sigma}_1} J + \int_{\tau} J = 0
\label{eq:stokes_expanded}
\end{equation}
The logic now is to take the limit as the compact regions $\tilde{\Sigma}_1$ and $\tilde{\Sigma}_2$ expand to fill their entire respective Cauchy surfaces. In this limit, the connecting surface $\tau$ recedes to the timelike cylinder at spatial infinity, $\tau_\infty$. Suppose that the integrals over the Cauchy surfaces remain well-defined and convergent in this limit, and critically, that the flux integral over the cylinder vanishes, i.e.,
\begin{equation}
     \lim_{\tau \to \tau_\infty} \int_\tau J = 0. \label{eq:vanishing_the_flux}
\end{equation}
If this condition holds, then Eq. \eqref{eq:stokes_expanded} implies that the symplectic structure is conserved
$$ \int_{\Sigma_2} J = \int_{\Sigma_1} J $$
This would mean that the integral of the symplectic current is independent of the choice of Cauchy surface, which is the necessary condition for a well-defined pre-symplectic structure $\Omega$. However the question of whether the asymptotic boundary conditions truly ensure the convergence of these integrals and the vanishing of the flux across $\tau_\infty$ is a delicate and crucial issue. If either of these properties were to fail, we would not obtain a consistent symplectic structure on our phase space $\Gamma$.

Therefore, the primary task for the remainder of our work is to analyze the asymptotic behavior of the symplectic current from Eq. \eqref{eq:SymplecticCurrentHalfShell_main_sec2} using the field expansions from Section 3. We must then identify boundary conditions (specifically, parity conditions) that guarantee the vanishing of this flux. This will ensure that our phase space is equipped with a consistent and conserved symplectic structure, providing a solid foundation for defining the charges associated with the BMS group.

Thus we need to show that the requirement \eqref{eq:vanishing_the_flux} is met. First note that $\lim_{\tau \to \tau_\infty} \int_\tau J = \lim_{\tau \to \tau_\infty} \int_\tau J_P + \lim_{\tau \to \tau_\infty} \int_\tau J_H.$ The contribution of the Holst term to the symplectic flux, $\int_{\tau_\infty} J_H$, vanishes for a straightforward geometric reason. From Eq. \eqref{eq:SymplecticCurrentHalfShell_main_sec2}, we see that the Holst current is an exact 3-form. By applying Stokes' theorem, the integral of this exact form over the 3-dimensional surface $\tau$ can be converted into an integral of the 2-form $K_H = \frac{1}{\kappa\beta}(\delta_{1}e^{I}\wedge\delta_{2}e_{I})$ over the boundary of $\tau$ which connects the two Cauchy surfaces $\Sigma_1$ and $\Sigma_2$ at spatial infinity. This boundary consists of two components: the 2-sphere at infinity on the future surface, $S_\infty(t_2)$, and the 2-sphere at infinity on the past surface, $S_\infty(t_1)$. Due to their orientation, the integral becomes:
$$\oint_{\partial\tau} K_H = \oint_{S_\infty(t_2)} K_H - \oint_{S_\infty(t_1)} K_H$$
Now, we invoke the physical requirement of stationarity at spatial infinity. This means we assume that the physical state of the isolated system, as seen from spatial infinity, does not change over time. Since the 2-form $K_H$ is constructed from the fields, its integral over the sphere at infinity must be independent of the time at which the slice is taken.
we invoke the intrinsic asymptotic behavior of the gravitational field at spatial infinity. While the bulk spacetime may be highly dynamical and contain gravitational radiation, these radiative modes (Bondi news) propagate towards null infinity ($\mathscr{I}$). As we approach spatial infinity ($i^0$) along spacelike Cauchy surfaces, the flux of gravitational radiation asymptotically vanishes. Consequently, the leading-order magnetic part of the boundary geometry becomes stationary, as rigorously established by Ashtekar and Hansen \cite{Ashtekar:1978zani}. Because the 2-form $K_H$ is constructed from these stationary leading-order fields, its integral over the sphere at infinity is invariant under time translation.

The two boundary integrals are equal and thus cancel each other exactly, leading to the conclusion that the flux of the Holst part is zero:
$$\int_{\tau_\infty} J_H = 0$$
This shows that the Holst term does not contribute to any potential leakage of the symplectic structure at spatial infinity. Therefore, to ensure a well-defined symplectic form, we only need to analyze the flux of the Palatini part, $\int_{\tau_\infty} J_P$.

\begin{equation}
    \int_{\tau} J_P = -\frac{1}{\kappa} \int_{\tau}  \text{Tr}[(\delta_{[1} \Sigma_{\sigma \mu}) (\delta_{2]} \omega_\nu)] \epsilon^{\sigma \mu \nu \lambda} \hat{x}_\lambda \; \sqrt{|\det(h)|} \; \rho^3 d\chi d^2\Omega_o
    \label{eq:epsilonTimesCurrent}
\end{equation}
where $d^2\Omega_o$ is the standard area element of the unit 2-sphere and  
$\sqrt{|\det(h)|}$ is the volume element on the unit 3-hyperboloid, with $h_{ij}$ being the metric on this surface.
The components of the integrand scale as $\mathcal{O}(\rho^{-3})$, while the volume element of the 3-surface $\tau$ scales as $\rho^3 d\chi d^2\Omega_o$. Therefore, the overall expression is finite in the limit $\rho \to \infty$. However, finiteness is not sufficient; for the symplectic structure to be well-defined, this flux integral must vanish.

Let's begin by expanding the expression $(\delta_{[1}{}^1e_{\mu}^{L})({\delta_{2]}}{}^2\omega_{\nu}^{IJ})$ and denote the relevant parts of the fields
\begin{itemize}
    \item From (\ref{eq:1e_derived_r}): ${}^1e_\mu^L(\Phi) = \sigma(\Phi) A_\mu^L(\Phi) + H_\mu^L(\Phi)$, where $A_\mu^L(\Phi) = \hat{x}_\mu(\Phi) \hat{x}^L(\Phi)$ and $H_\mu^L(\Phi)$ is shown in (\ref{eq:H_def_r})).
    \item From (\ref{eq:2omega_final_explicit_r}): ${}^2\omega_\nu^{IJ}(\Phi) = \Omega_{\sigma,\nu}^{IJ}(\Phi) + \Omega_{H,\nu}^{IJ}(\Phi)$, where
         $\Omega_{\sigma,\nu}^{IJ}(\Phi) = 2\left[(\mathcal{D}^{[J}\sigma)\hat{x}_\nu\hat{x}^{I]} + \sigma\delta_\nu^{[J}\hat{x}^{I]}\right]$ and
        $\Omega_{H,\nu}^{IJ}(\Phi) = 2\mathcal{D}^{[J}H_\nu^{I]}(\Phi) - 2H_\nu^{[I}(\Phi)\hat{x}^{J]}(\Phi)$
\end{itemize}
The variations $\delta_s$ (for $s=1,2$) act on the field coefficients $\sigma(\Phi)$ and ${}^1h_{AB}(\Phi)$ and the geometric quantities $\hat{x}_\mu, \hat{x}^I, \hat{e}^A_\mu, \hat{e}^{A I}$ and the angular derivative operator $\mathcal{D}^J$ are treated as background structures here (functions of $\Phi$ but not varied by $\delta_s$).
Thus, the variations are
\begin{itemize}
    \item $\delta_s {}^1e_\mu^L = (\delta_s\sigma) A_\mu^L + \delta_s H_\mu^L$ where $\delta_s H_\mu^L = \frac{1}{2} (\delta_s {}^1h_{AB}) \hat{e}^A_\mu \hat{e}^{BL}$,
    \item $\delta_s {}^2\omega_\nu^{IJ} = \delta_s \Omega_{\sigma,\nu}^{IJ} + \delta_s \Omega_{H,\nu}^{IJ}$, where
        $\delta_s \Omega_{\sigma,\nu}^{IJ} = 2\left[(\mathcal{D}^{[J}(\delta_s\sigma))\hat{x}_\nu\hat{x}^{I]} + (\delta_s\sigma)\delta_\nu^{[J}\hat{x}^{I]}\right]$ and 
        $\delta_s \Omega_{H,\nu}^{IJ} = 2\mathcal{D}^{[J}(\delta_s H_\nu^{I]}) - 2(\delta_s H_\nu^{[I)})\hat{x}^{J]}$
\end{itemize}
Let $e_1 = (\delta_1\sigma) A_\mu^L$, $e'_1 = \delta_1 H_\mu^L$. So $\delta_1 {}^1e_\mu^L = e_1 + e'_1$.
Let $\omega_1 = \delta_1 \Omega_{\sigma,\nu}^{IJ}$, $\omega'_1 = \delta_1 \Omega_{H,\nu}^{IJ}$. So $\delta_1 {}^2\omega_\nu^{IJ} = \omega_1 + \omega'_1$.
Similarly for $\delta_2$. The expression becomes
$$ \frac{1}{2} \left[ (e_1+e'_1)(\omega_2+\omega'_2) - (e_2+e'_2)(\omega_1+\omega'_1) \right] $$
Expanding this, we get four main groups of terms:
$$ \frac{1}{2} \left[ (e_1\omega_2 - e_2\omega_1) + (e_1\omega'_2 - e_2\omega'_1) + (e'_1\omega_2 - e'_2\omega_1) + (e'_1\omega'_2 - e'_2\omega'_1) \right] $$
So the full expression for $X_{\mu\nu}^{LIJ}$ can be decomposed into four types of terms based on their dependence on variations of $\sigma(\Phi)$ and $H_\mu^L(\Phi)$ as follows.
\\

\textbf{Term A: $(e_1\omega_2 - e_2\omega_1)$ (Purely $\sigma$-dependent parts)}
\begin{align*}
\text{Term A} &= ((\delta_{[1}\sigma) A_\mu^L) (\delta_{2]} \Omega_{\sigma,\nu}^{IJ})  \\
&=
2 A_\mu^L  (\delta_{[1}\sigma) \left[(\mathcal{D}^{[J}(\delta_{2]}\sigma))\hat{x}_\nu\hat{x}^{I]} + (\delta_{2]}\sigma)\delta_\nu^{[J}\hat{x}^{I]}\right] \\
&=
2 \hat{x}_\mu\hat{x}^L \hat{x}_\nu\hat{x}^{[I} \left( (\delta_{[1}\sigma) \mathcal{D}^{J]}(\delta_{2]}\sigma) \right)
\end{align*}

\textbf{Term B: $(e_1\omega'_2 - e_2\omega'_1)$ ($\sigma$-$H$ cross term)}
\begin{align*}
\text{Term B} &= ((\delta_{[1}\sigma) A_\mu^L) (\delta_{2]} \Omega_{H,\nu}^{IJ}) \\
&=
2 A_\mu^L (\delta_{[1}\sigma) \left(\mathcal{D}^{[J}(\delta_{2]} H_\nu^{I]}) - \delta_{2]} H_\nu^{[I}\hat{x}^{J]}\right) \\
&= 
2 \hat{x}_\mu \hat{x}^L (\delta_{[1}\sigma) \left(\mathcal{D}^{[J}(\delta_{2]} H_\nu^{I]}) - \delta_{2]} H_\nu^{[I}\hat{x}^{J]}\right)
\end{align*}

\textbf{Term C: $(e'_1\omega_2 - e'_2\omega_1)$ ($H$-$\sigma$ cross term)}
\begin{align*}
\text{Term C} &=  (\delta_{[1} H_\mu^L) (\delta_{2]} \Omega_{\sigma,\nu}^{IJ}) \\
&= 
2(\delta_{[1} H_\mu^L) \left[(\mathcal{D}^{[J}(\delta_{2]}\sigma))\hat{x}_\nu\hat{x}^{I]} + (\delta_{2]}\sigma)\delta_\nu^{[J}\hat{x}^{I]}\right] 
\end{align*}

\textbf{Term D: $(e'_1\omega'_2 - e'_2\omega'_1)$ (Purely $H$-dependent parts)}
\begin{align*}
\text{Term D} &=  (\delta_{[1} H_\mu^L) (\delta_{2]} \Omega_{H,\nu}^{IJ})\\
&= 2 (\delta_{[1} H_\mu^L) \left[\mathcal{D}^{[J}(\delta_{2]} H_\nu^{I]}) - \delta_{2]} H_\nu^{[I}\hat{x}^{J]}\right] 
\end{align*}
Therefore, we have 
$$(\delta_{[1}{}^1e_{\mu}^{L})({\delta_{2]}}{}^2\omega_{\nu}^{IJ}) = \text{Term A}_{\mu\nu}^{LIJ} + \text{Term B}_{\mu\nu}^{LIJ} + \text{Term C}_{\mu\nu}^{LIJ} + \text{Term D}_{\mu\nu}^{LIJ} $$

Plugging this result into Eq. \eqref{eq:epsilonTimesCurrent} and taking the limit, one gets
\begin{equation}
    \int_{\tau_\infty} J_P = -\frac{1}{\kappa} \int_{\tau_\infty}  \mathcal{F}(\Phi) \; \sqrt{|\det(h)|} \; d\chi d^2\Omega_o 
    \label{eq:SymlolycIntegralOfJP}
\end{equation}
where
\begin{align}
    \mathcal{F}(\Phi) &= \text{Tr}[(\delta_{[1} \Sigma_{\sigma \mu}) (\delta_{2]} \omega_\nu)] \epsilon^{\sigma \mu \nu \lambda} \hat{x}_\lambda
 \nonumber \\ 
    &= 
2 \epsilon^{\sigma \mu \nu \lambda} \hat{x}_\lambda \epsilon_{IJKL} \delta^K_\sigma 
(\delta_{[1} H_\mu^L) \left((\mathcal{D}^{[J}(\delta_{2]}\sigma))\hat{x}_\nu\hat{x}^{I]} + (\delta_{2]}\sigma)\delta_\nu^{[J}\hat{x}^{I]}\right) \nonumber \\
&\quad \; 
+2 \epsilon^{\sigma \mu \nu \lambda} \hat{x}_\lambda \epsilon_{IJKL} \delta^K_\sigma 
(\delta_{[1} H_\mu^L) \left(\mathcal{D}^{[J}(\delta_{2]} H_\nu^{I]}) - \delta_{2]} H_\nu^{[I}\hat{x}^{J]}\right)
\label{eq:flux_integrand_expanded}
\end{align}
in which we have used the fact that $\epsilon^{\sigma \mu\nu \lambda} \text{Term A}_{\mu\nu}^{LIJ} = 0$ since it is symmetric over its lower indices $\mu$ and $\nu$, and $\epsilon^{\sigma \mu \nu \lambda} \hat{x}_\lambda \text{Term B}_{\mu\nu}^{LIJ} = 0$ because the contraction of the symmetric product $\hat{x}_\mu \hat{x}_\lambda$ with the totally antisymmetric tensor $\epsilon^{\sigma \mu \nu \lambda}$ vanishes identically.

It is instructive to recall that in the frameworks of \cite{Ashtekar:2008jw, Corichi:2010aaa}, the condition ${}^1h_{AB} = -2\sigma h_{AB}$ was imposed to eliminate supertranslations. Under this restriction, the term $H_\mu^I$ is no longer independent of $\sigma$, and a direct calculation shows that the integrand \eqref{eq:flux_integrand_expanded} vanishes, leading to a well-defined symplectic structure for the Poincar\'e group.

Interestingly, while the full integrand \eqref{eq:flux_integrand_expanded} is significantly more complicated in the presence of supertranslations, it can still be shown to vanish identically. This result is not immediately obvious but follows from a careful application of the geometric properties of the fields. Specifically, it is a direct consequence of the tangentiality of $H_\mu^I$ and the identities governing its covariant derivatives with respect to the normal vector $\hat{x}_\mu$:
\begin{equation}
\hat{x}_I H^I_\mu = 0, \quad
\hat{x}_I \mathcal{D}^I H^J_\mu = 0, \quad 
\hat{x}_J \mathcal{D}^I H^J_\mu = -H^I_\mu
\end{equation}
This demonstrates that the inclusion of supertranslations does not lead to a leakage of the symplectic flux at spatial infinity, i.e. $\int_{\tau_\infty} J_P=0$. 

A crucial next step is to ensure that the total pre-symplectic structure on the phase space, given by the integral of the symplectic current $\int_\Sigma J(\delta_1, \delta_2)$, is well-defined. The finiteness of this integral is essential for a consistent Hamiltonian framework, yet it is not immediately guaranteed. A standard power-counting argument suggests a potential divergence. The symplectic current $J$ is a 3-form whose leading term, arising from the product of $\delta\Sigma \sim \mathcal{O}(\rho^{-1})$ and $\delta\omega \sim \mathcal{O}(\rho^{-2})$, falls off as $\mathcal{O}(\rho^{-3})$. Since the volume element on the hyperboloid $\Sigma$ behaves as $\rho^2 d\rho d^2\Phi$, the leading term in the integral scales as $\int (d\rho/\rho)$, indicating a potential logarithmic divergence.

It is instructive to clarify the relationship between the standard approach to spatial infinity and the hyperbolic coordinates used herein. The spatial infinity limit ($i^0$) is typically taken along a hypersurface of constant time, $t=t_0$. Using the coordinate transformation $t = \rho\sinh(\chi)$, this condition implies that as the spatial distance goes to infinity (and thus $\rho \to \infty$), the hyperbolic angle must approach zero ($\chi \to 0$) in such a way that the product $\rho\sinh(\chi)$ remains finite and equal to $t_0$. Therefore, the limit of approaching spatial infinity along a constant-time slice is described by the simultaneous limit $\rho \to \infty$ and $\chi \to 0$.

Following the calculations in this section, it can be straightforwardly shown that the pre-symplectic structure is given by
\begin{align}
\Omega (\delta_1, \delta_2) &= -\frac{1}{\kappa}\int_\Sigma J(\delta_1, \delta_2) = -\frac{1}{\kappa}\int_\Sigma \text{Tr}[(\delta_{[1} \Sigma_{\sigma \mu}) (\delta_{2]} \omega_\nu)] \epsilon^{\sigma \mu \nu \lambda} n_\lambda \rho^2 d\rho d^2\Omega_o+\frac{1}{\kappa\beta} \int_\Sigma d(\delta_{1}e^{I}\wedge\delta_{2}e_{I})
 \nonumber \\ 
    &=
-\frac{1}{\kappa}\int \frac{ d\rho}{\rho} \int_{S^2} d^2\Omega_o \left[2 \epsilon^{\sigma \mu \nu \lambda} n_\lambda \epsilon_{IJKL} \delta^K_\sigma \hat{x}_\mu \hat{x}^L(\delta_{[1}\sigma) \left(\mathcal{D}^{[J}(\delta_{2]} H_\nu^{I]}) - \delta_{2]} H_\nu^{[I}\hat{x}^{J]}\right)\right. \nonumber\\
&\qquad \qquad \qquad \qquad \qquad \;
 +2 \epsilon^{\sigma \mu \nu \lambda} n_\lambda \epsilon_{IJKL} \delta^K_\sigma 
(\delta_{[1} H_\mu^L) \left((\mathcal{D}^{[J}(\delta_{2]}\sigma))\hat{x}_\nu\hat{x}^{I]} + (\delta_{2]}\sigma)\delta_\nu^{[J}\hat{x}^{I]}\right) \nonumber \\
&\qquad \qquad \qquad \qquad \qquad \;
\left.+2 \epsilon^{\sigma \mu \nu \lambda} n_\lambda \epsilon_{IJKL} \delta^K_\sigma 
(\delta_{[1} H_\mu^L) \left(\mathcal{D}^{[J}(\delta_{2]} H_\nu^{I]}) - \delta_{2]} H_\nu^{[I}\hat{x}^{J]}\right)\right] + \mathcal{O}(\rho^{-1})
\end{align}
where $n_\mu$ is the unit normal to the hypersurface $\Sigma$, given by $n_\mu =(1,0,0,0)$ in the asymptotic Cartesian coordinates.

The presence of the $\int (d\rho/\rho)$ term indicates a logarithmic divergence. Unlike the symplectic flux at spatial infinity discussed in \eqref{eq:SymlolycIntegralOfJP} and \eqref{eq:flux_integrand_expanded}, the integrand here is not identically zero. Consequently, the pre-symplectic form contains a divergent piece that must be handled to ensure a well-defined Hamiltonian framework.
After a considerable amount of algebraic simplification, the divergent part of the pre-symplectic structure can be isolated and expressed as
\begin{align}
\Omega (\delta_1, \delta_2)_{\text{div}} =
-\frac{2}{\kappa}\int \frac{ d\rho}{\rho} \int_{S^2} d^2\Omega_o 
&\lim_{\chi\to 0}\left[
(\delta_{[1} H_I^t)(\mathcal{D}^{I}(\delta_{2]}\sigma)- (\delta_{[1} H_L^L)(\mathcal{D}^{t}(\delta_{2]}\sigma)\right.\nonumber\\
&\quad \quad
- (\delta_{[1}\sigma) \mathcal{D}^t(\delta_{2]} H_I^I) 
+(\delta_{[1}\sigma) \mathcal{D}^{J}(\delta_{2]} H_J^t)
\nonumber\\ 
&\quad \quad
+(\delta_{[1} H_I^t) \mathcal{D}^{I}(\delta_{2]} H_J^{J}) - (\delta_{[1} H_I^L) \mathcal{D}^{I}(\delta_{2]} H_L^t)\nonumber\\
&\quad \quad
-(\delta_{[1} H_J^t) \mathcal{D}^{I}(\delta_{2]} H_I^{J}) + (\delta_{[1} H_J^L) \mathcal{D}^t(\delta_{2]} H_L^{J})\nonumber\\
&\quad \quad
\left.+ (\delta_{[1} H_L^L) \mathcal{D}^{I}(\delta_{2]} H_I^t) - (\delta_{[1} H_L^L) \mathcal{D}^t(\delta_{2]} H_J^{J})\right]
\label{eq:Symplectic_div_2}
\end{align}
where we have used the vanishing of the time component of the radial vector, $\hat{x}^t \to 0$, in the spatial infinity limit  $\chi \to 0 $ on $\Sigma$. 
It is worth noting that while the limit and the covariant derivative operator do not commute in general, the equality $\lim_{\chi \to 0}\mathcal{D}^I H^J_\mu = \mathcal{D}^I (\lim_{\chi \to 0} H^J_\mu)$ holds in this specific context. A full proof requires expanding the covariant derivative via the product rule, $\mathcal{D}^I(H^J_\mu) = \mathcal{D}^I(\frac{1}{2}{}^1h_{AB}\hat{e}^A_\mu \hat{e}^{BJ})$. The key insight is that the additional terms generated by the product rule, which involve derivatives of the basis vectors (i.e., the connection coefficients of the hyperboloid), vanish in the $\chi \to 0$ limit. Specifically, connection components such as $\Gamma^{\bar{A}}_{\chi\bar{B}} = \delta^{\bar{A}}_{\bar{B}}\tanh(\chi)$ go to zero. Consequently, the derivative operator $\mathcal{D}^I$ effectively reduces to the covariant derivative operator on the asymptotic 2-sphere, and the identity holds.
Therefore, we can evaluate the relevant expressions by first taking the $\chi \to 0$ limit of the fields inside the covariant derivative operator. This allows us to replace the full hyperboloidal field $H^I_\mu$ with its asymptotic limit on the 2-sphere at spatial infinity. The components of $H^I_\mu$ in this limit, where the inner indices $i, j, k, \dots$ are understood to be spatial,  are given by
\begin{align}
    \lim_{\chi \to 0} H^0_t &= \frac{1}{2} {}^1\bar{h}_{\chi \chi} \\
    \lim_{\chi \to 0} H^i_t &= -\frac{1}{2}{}^1\bar{h}_{\chi \bar{A}} \hat{e}^{\bar{A} i} \\
    \lim_{\chi \to 0} H^0_a &= -\frac{1}{2}{}^1\bar{h}_{\chi \bar{A}} \hat{e}^{\bar{A}}_a =: -\bar{\lambda}_i \delta^i_a \\
    \lim_{\chi \to 0} H^i_a &= \frac{1}{2}\left({}^1\bar{h}_{\bar{A}\bar{B}} \hat{e}^{\bar{A}}_a \hat{e}^{\bar{B} i}\right) =: \bar{H}^i_j \delta^j_a
\end{align}
where we denote quantities evaluated in the $\chi\to 0$ limit with a bar, e.g., $\lim_{\chi\to 0} {}^1h_{\chi\bar{A}} =: {}^1\bar{h}_{\chi\bar{A}}$ and we have introduced the fields $\bar{\lambda}_i$ and $\bar{H}^i_j$ which live on the asymptotic 2-sphere. Substituting these limiting forms into the expression \eqref{eq:Symplectic_div_2} and simplifying, we have
\begin{align}
\Omega (\delta_1, \delta_2)_{\text{div}} =
-\frac{2}{\kappa}\int \frac{ d\rho}{\rho} \int_{S^2} d^2\Omega_o 
&\left[
-(\delta_{[1} \bar{\lambda}_i)(\mathcal{D}^i(\delta_{2]}\sigma) - (\delta_{[1} \bar{H}_i^i)(\mathcal{D}^t(\delta_{2]}\sigma)\right.\nonumber\\
&\quad 
- (\delta_{[1}\sigma) \mathcal{D}^t(\delta_{2]} \bar{H}_i^i) -(\delta_{[1}\sigma) \mathcal{D}^i(\delta_{2]} \bar{\lambda}_i)\nonumber\\
&\quad
+(\delta_{[1}\bar{\lambda}_i) \mathcal{D}^i(\delta_{2]} \bar{H}_j^j) 
- (\delta_{[1} \bar{H}_i^j) \mathcal{D}^i(\delta_{2]} \bar{\lambda}_j) \nonumber \\ 
&\quad
-(\delta_{[1} \bar{\lambda}_j) \mathcal{D}^{i}(\delta_{2]} \bar{H}_i^j)
+ (\delta_{[1} \bar{H}_i^j) \mathcal{D}^t(\delta_{2]} \bar{H}_j^i)\nonumber \\
&\quad
\left.+ (\delta_{[1} \bar{H}_j^j) \mathcal{D}^i(\delta_{2]} \bar{\lambda}_i)  
- (\delta_{[1} \bar{H}_i^i) \mathcal{D}^t(\delta_{2]} \bar{H}_j^j) \right]
\label{eq:Symplectic_div_3}
\end{align}
A strategy to eliminate this divergence is to assign parities to the various fields such that the integrand of the $S^2$
integral becomes an odd function, causing the integral to vanish. The key challenge is to impose these parity conditions in a physically and mathematically consistent manner. Specifically, the assignments must not cause the charges of the asymptotic symmetries to vanish or become non-integrable. Furthermore, the parity conditions themselves must be invariant under the action of the asymptotic symmetry group (Lorentz transformations and supertranslations). With these considerations in mind, we will introduce the required parity conditions in the following section.

\section{\textsf{Parity Conditions for a Well-Defined Pre-Symplectic Structure}}\label{Section5}
First, it is worth noting that in \cite{Ashtekar:2008jw, Corichi:2010aaa} the logarithmic ambiguities were eliminated by imposing the condition that $\sigma(\Phi)$ is even under the antipodal map on the hyperboloid 
\begin{equation}\label{antipodal map}
    (\chi, \theta, \phi) \to (-\chi, \pi -\theta, \pi + \phi). 
\end{equation}
We adopt this condition as well, as our primary focus is to include supertranslations into the Hamiltonian framework, and not the logarithmic terms.
It is important, then, to ensure that this parity choice is consistent with the asymptotic symmetries. Under an infinitesimal transformation generated by a vector field $\xi^\mu$, the scalar field $\sigma$ transforms by
\begin{equation}
    \delta_\xi \sigma = \mathcal{L}_{X} \sigma
\end{equation}
A direct analysis of the transformation properties of the Killing vectors $X^A$ (Recall \eqref{definition of X}) and the gradient of $\sigma$ under the antipodal map shows that if $\sigma$ is an even function, its Lie derivative $\mathcal{L}_{X} \sigma$ is also an even function. This confirms that our even parity condition on $\sigma$ is preserved under the full asymptotic symmetry group.

Another boundary condition consistently used throughout our framework is the vanishing of the leading order $g_{\rho A}$ components, i.e., ${}^1h_{\rho A}=0$ in the metric expansion \eqref{eq:GeneralMetricHyperbolic_r_revised}. It is crucial to examine the conditions under which this gauge choice is preserved by the asymptotic symmetry group generated by $\xi$. From the leading order term ($\mathcal{O}(\rho^0)$) of the Lie derivative $\mathcal{L}_\xi g_{\rho A}$
\begin{equation}
    \delta_\xi {}^1h_{\rho A} = \nabla_A (S + {}^1h_{\rho B}  X^B) - S_A + \mathcal{L}_X {}^1h_{\rho A}
\end{equation}
we find that preserving ${}^1h_{\rho A}=0$ imposes the following condition on the tangential components of the supertranslation generator:
\begin{equation} \label{eq:gauge_preservation_condition}
    S_A = \nabla_A S
\end{equation}
Recall that $S_A := S^\mu \hat{e}_{A\mu}$ are the components of the supertranslation in the tangential basis, while $S := S^\mu \hat{x}_\mu$ is the radial component, and $\nabla_A$ is the covariant derivative compatible with the hyperboloid metric $h_{AB}$.
By defining the spatial radial component of the supertranslation generator as $W := S^a \hat{x}_{a, \text{spatial}}$\footnote{$\hat{x}_{a, \text{spatial}}$ are the components of the standard 3D spatial unit radial .}, it can be shown that the $\chi$ component of $S_A$ and the radial component $S$ are explicitly given by
\begin{equation}\label{expression S}
    S_\chi= -S^t \cosh\chi + W \sinh \chi, \qquad S = -S^t \sinh \chi + W \cosh \chi
\end{equation}
Assuming the standard condition that the supertranslation generator $S^\mu$ is independent of $\chi$, i.e., $\partial_\chi S^\mu = 0$, it can be readily verified that for the component $A=\chi$, relation \eqref{eq:gauge_preservation_condition} reduces to an identity which is automatically satisfied\footnote{Explicitly, the condition becomes $\nabla_\chi S - S_\chi = \partial_\chi S - S_\chi = (\partial_\chi W)\cosh\chi - (\partial_\chi S^t)\sinh\chi = 0$, which holds if $S^\mu$ is $\chi$-independent.}.

We now turn to the parity assignments for the components of ${}^1h_{AB}$, which play a crucial role in eliminating the divergent part of the pre-symplectic structure. Let us first consider the component ${}^1h_{\chi\chi}$. Notably, this component does not appear in the expression for $\Omega (\delta_1, \delta_2)_{\text{div}}$. Consequently, the requirement of removing the divergence imposes no restriction on its parity. Thus, ${}^1h_{\chi\chi}$ can, in general, possess both even and odd parts under the antipodal map
\begin{equation}
    {}^1h_{\chi\chi} = ({}^1h_{\chi\chi})_{\text{odd}} + ({}^1h_{\chi\chi})_{\text{even}}
\end{equation}
This decomposition is preserved by the asymptotic symmetries. As shown in Appendix \ref{app:field_transformations}, the transformation of ${}^1h_{AB}$ under an infinitesimal symmetry transformation generated by $\xi$ is given by\footnote{One can explicitly show that $\nabla_\chi S_\chi - S =0$.}
\begin{equation}
    \delta_\xi {}^1h_{\chi \chi} = \mathcal{L}_X {}^1h_{\chi \chi}
\end{equation}
Since the Lie derivative with respect to $X$ preserves the parity properties of tensor fields on the hyperboloid, this transformation law separately maps the even and odd parts of ${}^1h_{\chi\chi}$ to themselves. 

To address the divergent part of the pre-symplectic structure, it is convenient to introduce new variables defined on the asymptotic 2-sphere ($\chi=0$). We define the trace-adjusted spatial part of the first-order metric perturbation as
\begin{equation}
{}^1\bar{k}_{\bar{A}\bar{B}} := \frac{1}{2}{}^1\bar{h}_{\bar{A}\bar{B}} + \sigma \bar{h}_{\bar{A}\bar{B}}, \qquad {}^1\bar{k}_j^i := {}^1\bar{k}_{\bar{A}\bar{B}} \hat{e}^{\bar{A}}_a \delta^a_j \hat{e}^{\bar{B} i}
\end{equation}
where $\bar{h}_{\bar{A}\bar{B}} = \lim_{\chi\to 0} h_{\bar{A}\bar{B}}$ is the metric on the unit 2-sphere. From this definition, the trace ${}^1\bar{k}$ follows as:
\begin{equation}
{}^1\bar{k} = \bar{h}^{\bar{A}\bar{B}} {}^1\bar{k}_{\bar{A}\bar{B}} = \frac{1}{2}{}^1\bar{h}_{\bar{A}\bar{B}} \bar{h}^{\bar{A}\bar{B}} + 2\sigma
\end{equation}
In terms of these variables and the previously defined $\bar{\lambda}_i = \lim_{\chi\to 0} \lambda_i$, the divergent part of the pre-symplectic structure takes the form
\begin{align}
&\Omega (\delta_1, \delta_2)_{\text{div}} =\nonumber\\
&
-\frac{2}{\kappa}\int \frac{ d\rho}{\rho} \int_{S^2} d^2\Omega_o \left[ 2(\delta_{[1}\sigma) \mathcal{D}^t(\delta_{2]}\sigma) 
-(\delta_{[1} \bar{\lambda}_i) \mathcal{D}^i(\delta_{2]} {}^1\bar{k}) 
+ (\delta_{[1} {}^1\bar{k}_i^j) \mathcal{D}^i(\delta_{2]} \bar{\lambda}_j)
+(\delta_{[1} \bar{\lambda}_j) \mathcal{D}^{i}(\delta_{2]} {}^1\bar{k}_i^j)\right.
\nonumber\\ 
 &\qquad \qquad \qquad \qquad \qquad \; \; \left.+ (\delta_{[1} {}^1\bar{k}_i^j) \mathcal{D}^t(\delta_{2]}{}^1\bar{k}_j^i) 
- (\delta_{[1} {}^1\bar{k}) \mathcal{D}^i(\delta_{2]} \bar{\lambda}_i)
- (\delta_{[1} {}^1\bar{k}) \mathcal{D}^t(\delta_{2]} {}^1\bar{k}) \right] \label{eq:Omega_div_rewritten}
\end{align}
It is now evident that the divergence can be eliminated by assigning odd parity to the remaining fields
\begin{equation} \label{eq:parity_assignments}
    \bar{\lambda}_i = \text{odd}, \qquad {}^1\bar{k}_j^i = \text{odd} 
\end{equation}
then each term within the square brackets in \eqref{eq:Omega_div_rewritten} becomes an odd function. The required odd parities for $\bar{\lambda}_i$ and ${}^1\bar{k}_{\bar{A}\bar{B}}$ in \eqref{eq:parity_assignments} can be derived from the following fundamental parity assignments
\begin{equation} \label{eq:fundamental_h_parities}
\begin{alignedat}{2}
    {}^1h_{\chi \theta} &= \text{odd}, &\qquad {}^1h_{\chi \phi} &= \text{even} \\
    {}^1k_{\theta \theta} &= \text{odd}, &\qquad {}^1k_{\theta \phi} &= \text{even}, &\qquad {}^1k_{\phi \phi} &= \text{odd}
\end{alignedat}
\end{equation}
Now we need to check whether these assignments are preserved under the action of the asymptotic symmetry group.
From Appendix \ref{app:field_transformations}, we know their transformation rules 
\begin{align}
    \delta_\xi {}^1h_{\chi \bar{A}} &= \nabla_\chi S_{\bar{A}} + \nabla_{\bar{A}} S_\chi + \mathcal{L}_X {}^1h_{\chi \bar{A}} \label{eq:delta_h_chiA} \\
    \delta_\xi {}^1k_{\bar{A} \bar{B}} &= \frac{1}{2}(\nabla_{\bar{A}} S_{\bar{B}} + \nabla_{\bar{B}} S_{\bar{A}}) + S h_{\bar{A}\bar{B}} + \mathcal{L}_X {}^1 k_{\bar{A}\bar{B}} \label{eq:delta_h_AB}
\end{align}
To obtain the parity preservation, we assign definite parities to the fundamental supertranslation generators, namely the time translation $S^t$ and the spatial radial component $W$
\begin{equation} \label{eq:ST_parities}
    S^t = \text{even}, \qquad W = \text{odd}
\end{equation}
Recalling the expressions \eqref{expression S}, and using the condition \eqref{eq:gauge_preservation_condition} along with the properties of the covariant derivative under the antipodal map, we deduce the parities of the components $S$ and $S_A$ as
\begin{equation} \label{eq:S_component_parities}
    S = \text{odd}, \qquad S_\chi = \text{even}, \qquad S_\theta = \text{even}, \qquad S_\phi = \text{odd}
\end{equation}
Considering these parities \eqref{eq:ST_parities} \eqref{eq:S_component_parities} and the assumed parities \eqref{eq:fundamental_h_parities} into the transformation laws \eqref{eq:delta_h_chiA} and \eqref{eq:delta_h_AB}, it can be readily verified that the variations $\delta_\xi {}^1h_{\chi \bar{A}}$ and $\delta_\xi {}^1k_{\bar{A} \bar{B}}$ indeed possess the same parities as the original fields. Thus, our parity assignments required for the finiteness of the pre-symplectic structure are consistent with the asymptotic symmetries.

These conditions ensure that the pre-symplectic form 
\begin{equation}
    \Omega(\delta_{1},\delta_{2})= \int_\Sigma J(\delta_{1},\delta_{2})=-\frac{1}{2\kappa}\int_\Sigma \left[\delta_{1}\Sigma_{IJ}\wedge\delta_{2}\omega^{IJ}-\delta_{2}\Sigma_{IJ}\wedge\delta_{1}\omega^{IJ}\right]
    +\frac{1}{\kappa\beta} \int_{\partial \Sigma} \delta_{1}e^{I}\wedge\delta_{2}e_{I}
    \label{eq:Pre-Symplectic-Form_Palatini_Holst}
\end{equation}
is well-defined, providing a consistent starting point for defining generators of the asymptotic symmetry group in the next section. The pre-symplectic structure \eqref{eq:Pre-Symplectic-Form_Palatini_Holst} is thus composed of two distinct pieces: the familiar Palatini term, integrated over the bulk of the Cauchy surface $\Sigma$, and a surface term at infinity, $\partial \Sigma$, arising purely from the Holst part of the action which is generally non-zero. This structure elegantly reflects the nature of the Holst modification itself: while the Holst term in the action does not alter the local equations of motion, its contribution to the pre-symplectic structure manifests precisely as a boundary term, hinting at its potential influence on global quantities such as the conserved charges derived from $\Omega$. %
While the true physical symplectic form is formally obtained by quotienting this pre-symplectic structure by the action of gauge transformations, this step is not required for what we are seeking in this work. 

The crucial final step is to verify that the imposed parity conditions, while successfully eliminating the logarithmic divergence in $\Omega$, satisfy the essential ``physical'' requirements. That is, they must ensure that the resulting charges associated with the asymptotic symmetries are not only finite and integrable but also remain non-trivial, particularly for the supertranslations. Demonstrating that these conditions are met is the task of the next section, where we explicitly compute the charges.

\section{\textsf{Asymptotic Symmetries and Their Generators}}
\label{sec:charges}

Having established the asymptotic structure of the fields and the parity conditions required for a well-defined pre-symplectic structure in Section \ref{Section5}, we now proceed to the derivation of the conserved charges associated with the asymptotic symmetries. We employ the covariant phase space formalism, where the generator of a symmetry $\xi^\mu$ is identified as the Hamiltonian $H_\xi$ whose variation yields the contraction of the vector field with the symplectic form:
\begin{equation}
    \delta H_\xi = \Omega(\delta, \delta_\xi)
\end{equation}
The condition for the existence of such a generator is that the 1-form $X_\xi(\delta) := \Omega(\delta, \delta_\xi)$ be closed on the phase space, which corresponds to the integrability of the charge.

As derived in Eq. (2.6), the symplectic potential of the Holst action decomposes linearly into a Palatini part and a Holst part. Consequently, the total pre-symplectic form and the resulting charges also split into two distinct contributions:
\begin{equation}
    H_\xi = H_\xi^{P} + H_\xi^{H}
\end{equation}
To rigorously analyze the physical implications of the Holst term and the consistency of our boundary conditions, we will compute these contributions separately. In Section \ref{subsec:palatini_charges}, we calculate the generators arising from the standard Palatini symplectic structure. We will demonstrate that the parity conditions imposed to remove the symplectic flux divergence also render the Palatini charges finite and integrable, recovering the full set of BMS charges including supertranslations. In Section \ref{subsec:holst_charges}, we isolate the contribution of the Holst term to determine whether it modifies the physical values of the charges or acts purely as a boundary gauge artifact.

To explicitly evaluate the surface integrals for the charges in the following subsections, it is convenient to decompose the asymptotic symmetry vector field $\xi^\sigma$ into components tangential and orthogonal to the asymptotic hyperboloids. Here we need to recall from section \ref{sec3.2} that using the orthonormal basis $\{ \hat{x}^\sigma, \hat{e}^\sigma_A \}$, the generator takes the following form near spatial infinity
\begin{equation}
    \xi^\sigma = \rho X^A(\Phi) \hat{e}^\sigma_A + S(\Phi) \hat{x}^\sigma + S^A(\Phi) \hat{e}^\sigma_A
    \label{eq:xi_decomposition}
\end{equation}
where the components are identified as follows:
\begin{itemize}
    \item The term $\rho X^A \hat{e}^\sigma_A$ corresponds to the \textbf{Lorentz transformations} (boosts and rotations). Here, $X^A(\Phi)$ is a Killing vector field on the unit hyperboloid, satisfying $\mathcal{L}_X h_{AB} = 0$. The linear growth in $\rho$ is characteristic of rigid rotations and boosts at spatial infinity.
    \item The term $S(\Phi) \hat{x}^\sigma$ represents the radial component of the \textbf{supertranslations}. The function $S(\Phi)$ is the arbitrary parameter characterizing the angle-dependent translations.
    \item The term $S^A(\Phi) \hat{e}^\sigma_A$ is the tangential part of the supertranslations. As derived in Section \ref{Section5} (equation \eqref{eq:gauge_preservation_condition}), the preservation of the gauge condition ${}^1 h_{\rho A}=0$ constrains this component to be the gradient of the radial part, i.e., $S^A = \nabla^A S$.
\end{itemize}
With this decomposition, we can now systematically compute the contributions of the Palatini and Holst terms to the conserved charges.

\subsection{\textsf{Generators from the Palatini Term}}
\label{subsec:palatini_charges}

We begin by computing the contribution of the Palatini term to the asymptotic charges. The variation of the Hamiltonian generator associated with an asymptotic symmetry vector field $\xi^\mu$ is given by the contraction of the vector field with the symplectic form. For the Palatini action, applying the standard covariant phase space identities and Stokes' theorem, this variation reduces to a surface integral over the 2-sphere at spatial infinity ($S_\infty$):
\begin{equation}
    \delta H_{\xi}^{P} = \Omega_{P}(\delta, \delta_\xi) = \int_{\Sigma} \text{d} \left( i_\xi \Theta_P(\delta) - i_\delta \Theta_P(\xi) \right)
\end{equation}
Evaluated on the solution space where the torsion vanishes, this yields the standard expression:
\begin{equation}
    \delta H_{\xi}^{P} = -\frac{1}{2\kappa} \oint_{S_{\infty}} \text{Tr} \left[ (i_{\xi} \omega) \wedge \delta \Sigma - (i_{\xi} \Sigma) \wedge \delta \omega \right]
\end{equation}
where $i_{\xi}$ denotes the interior product with the vector field $\xi^\mu$. 

A naive power-counting argument suggests that this integral contains linearly divergent terms of order $\mathcal{O}(\rho)$. Specifically, the term involving $(i_\xi \Sigma) \wedge \delta \omega$ couples the Lorentz part of $\xi$ (which scales as $\rho$) with the leading order area element $\Sigma$ ($\sim \rho^2$) and the variation of the connection ($\sim \rho^{-2}$), potentially leading to a divergence.
Explicitly, the divergent part of the variation is given by
\begin{align}
X_\xi(\delta)=& \frac{2}{\kappa}  \rho \oint_{S_\infty} b (\delta \sigma) \nonumber \\
&
-\frac{1}{4\kappa}  \oint_{S_\infty} 
\left[b \left(2 \sigma  (\delta {}^1h_{\bar{A}\bar{B}})  h^{\bar{A} \bar{B}} -2   (\delta \sigma){}^1h_{\bar{A} \bar{B}} h^{\bar{A}\bar{B}}
+4 (\delta \sigma) {}^1h_{\chi \chi}\right. \right.\nonumber\\
& \qquad \qquad \qquad \qquad
\left. \left.-4 (\delta {}^3\omega^{IJ}_J) \hat{x}_I
-4 (\delta {}^3\omega^{IJ}_\nu)  \hat{e}^\nu_\chi \hat{e}^\chi_I \hat{x}_J  \right)\right.
\nonumber\\
&\;\;\;\;\;\;\; \;\;\;\;\;\;\;\;\;\;\;
+ X^{\bar{A}}\left( 2 \sigma  (\delta {}^1h_{\bar{A}\chi}) 
+2  (\delta \sigma){}^1h_{\bar{A}\chi} - 4  (\delta {}^3\omega^{IJ}_\nu) \hat{e}^\nu_{\bar{A}} \hat{e}^\chi_I \hat{x}_J\right)
\nonumber\\
&\;\;\;\;\;\;\; \;\;\;\;\;\;\;\;\;\;\;
\left.-4 S\; \hat{e}^\chi_J (\mathcal{D}^J (\delta H) - \mathcal{D}^I (\delta H^J_I))
-8 S^\chi (\delta \sigma) \right] +\mathcal{O}(\rho^{-1})
\end{align}
where we denoted the boost parameter by $b$, i.e., 
\begin{equation}
    b:= \lim_{\chi\to 0 } X^\chi
\end{equation}
that is an odd parameter.
The leading $\mathcal{O}(\rho)$ term in $X_\xi(\delta)$ must vanish for the charge to be finite, a condition which is satisfied by the imposed parity conditions under which the integral of the odd integrand $b(\delta\sigma)$ over the 2-sphere vanishes.

To extract the conserved charges, we decompose the variation $X_\xi(\delta)$ into two distinct contributions: an integrable part, where the variation $\delta$ can be factored out of the integral, and the remaining terms that initially appear non-integrable. By regrouping the terms derived from the Palatini symplectic structure, we obtain:

\begin{align}
X_\xi(\delta) =&
-\frac{1}{4\kappa} \delta \oint_{S_\infty} 
\left[b \left(
2 \sigma  ({}^1h_{\bar{A}\bar{B}})  h^{\bar{A} \bar{B}}
- 4 ({}^3\omega^{IJ}_J) \hat{x}_I
-4 ( {}^3\omega^{IJ}_\nu)  \hat{e}^\nu_\chi \hat{e}^\chi_I \hat{x}_J  \right) \right. \nonumber\\
& \left. \qquad \qquad
+ X^{\bar{A}}\left( 2 \sigma  \;{}^1h_{\bar{A}\chi}
- 4  ( {}^3\omega^{IJ}_\nu) \hat{e}^\nu_{\bar{A}} \hat{e}^\chi_I \hat{x}_J\right) -4 S\; \hat{e}^\chi_J (\mathcal{D}^J  H - \mathcal{D}^I  H^J_I) \right]\nonumber\\
&
-\frac{1}{4\kappa}  \oint_{S_\infty} 
\left[b \left(4   (\delta \sigma) {}^1h_{\chi \chi}  -4   (\delta \sigma){}^1h_{\bar{A} \bar{B}} h^{\bar{A}\bar{B}}
 \right)
-8 S^\chi (\delta \sigma) \right] +\mathcal{O}(\rho^{-1})
\label{eq:palatini_charge_split}
\end{align}

To proceed with the evaluation of these integrals at spatial infinity, we must consider the behavior of the symmetry generators in the limit where the hyperbolic angle vanishes ($\chi \to 0$). In this limit, the radial ($S$) and tangential ($S^\chi$) components of the supertranslation vector field relate to the standard time translation parameter $S^t$ and the spatial radial translation parameter $W$ according to
$\lim_{\chi\to 0} S = W,
\lim_{\chi\to 0} S^\chi = S^t$
Substituting these limits into the expression \eqref{eq:palatini_charge_split} allows us to determine the final form of the asymptotic charges.

To eliminate the non-integrable terms appearing in the second integral of \eqref{eq:palatini_charge_split} and obtain a well-defined Hamiltonian, we propose a redefinition of the asymptotic time translation parameter $S^t$. By inspecting the structure of the remaining terms, we define $S^t$ as follows:
\begin{equation}
    S^t := \frac{b}{2} ({}^1\bar{h}_{\chi \chi})_{\text{odd}} - b \, {}^1\bar{k} + T(\Phi)
    \label{eq:St_redefinition}
\end{equation}
where $T(\Phi)$ is an arbitrary \textit{even} function of the angles.
By promoting the symmetry parameter $S^t$ to be field-dependent, the corresponding canonical transformations are modified. This necessitates the use of a modified Lie algebroid bracket, analogous to the Barnich-Troessaert framework, to ensure the integrability and closure of the asymptotic algebra.
It is crucial to verify that this ansatz is consistent with the parity conditions established in \eqref{eq:ST_parities}. Let us examine the parity of each term in \eqref{eq:St_redefinition}:
    The term $({}^1\bar{h}_{\chi \chi})_{\text{odd}}$ is explicitly odd and the boost parameter $b$ is also odd, and therefore the product $b \, ({}^1\bar{h}_{\chi \chi})_{\text{odd}}$ is even.
     The trace term ${}^1\bar{k}$ was defined to be odd (\eqref{eq:fundamental_h_parities}). Consequently, the product $-b \, {}^1\bar{k}$ is also even.
    And finally we have explicitly chosen the arbitrary function $T(\Phi)$ to be even.
Thus, the sum of these terms yields an even $S^t$, ensuring that the redefinition \eqref{eq:St_redefinition} is fully compatible with the required parity conditions. Substituting this expression back into the charge variation cancels the obstructive terms, rendering the charge integrable.
\begin{align}
X_\xi(\delta) =&
-\frac{1}{4\kappa} \delta \oint_{S_\infty} 
\left[-8 T \sigma + b \left(
2 \sigma  ({}^1h_{\bar{A}\bar{B}})  h^{\bar{A} \bar{B}}
- 4 ({}^3\omega^{IJ}_J) \hat{x}_I
-4 ( {}^3\omega^{IJ}_\nu)  \hat{e}^\nu_\chi \hat{e}^\chi_I \hat{x}_J  \right) \right. \nonumber\\
& \left. \qquad \qquad \qquad
+ X^{\bar{A}}\left( 2 \sigma  \;{}^1h_{\bar{A}\chi}
- 4  ( {}^3\omega^{IJ}_\nu) \hat{e}^\nu_{\bar{A}} \hat{e}^\chi_I \hat{x}_J\right) -4 W\; \hat{e}^\chi_J (\mathcal{D}^J  H - \mathcal{D}^I  H^J_I)  \right]
\label{eq:palatini_charge_split_II}
\end{align}

Based on the integrable expression \eqref{eq:palatini_charge_split_II}, we can now isolate the generator corresponding to the supertranslation sector:
\begin{equation}
    Q_{\text{Supertranslation}} =  \frac{1}{\kappa} \oint_{S_\infty} 
    \left[2 \;T(\Phi) \sigma(\Phi)  + W(\Phi)\; \hat{e}^\chi_J (\mathcal{D}^J  H - \mathcal{D}^I  H^J_I)  \right]
    \label{eq:supertranslation_charge}
\end{equation}
This charge is manifestly non-vanishing. Since we have explicitly defined $T(\Phi)$ as an \textit{even} function and the mass aspect $\sigma(\Phi)$ satisfies an even parity condition, the product $T\sigma$ is an even function on the sphere. Consequently, its integral over $S^2$ is non-zero. Similarly, the spatial supertranslation parameter $W$ is odd, and it couples to field components with compatible parity to yield a non-trivial contribution.

It is instructive to compare this result with standard treatments in the literature, such as the framework developed in \cite{Ashtekar:2008jw}. In those approaches, the specific parity conditions imposed to ensure the finiteness of the symplectic structure often necessitate that the non-constant part of the supertranslation parameter $S^t$ be \textit{odd} to preserve the boundary conditions. Under such a restriction, the leading term in the charge integral would involve the product of an odd parameter with the even mass aspect $\sigma$, causing the integral to vanish identically. This effectively eliminates supertranslations as physical charges. In contrast, our formulation, through the redefinition \eqref{eq:St_redefinition} and the modified parity assignments, successfully regularizes the symplectic structure while retaining $S^t$ as an even function, thereby recovering the full, non-trivial supertranslation sector at spatial infinity.

\subsection{\textsf{Contribution of the Holst Term}}\label{subsec:holst_charges}
Before proceeding to the explicit calculation of the Holst charge, it is crucial to address a prevalent argument in the literature, notably in ~\cite{Corichi:2010aaa}, which suggests that the surface term arising from the Holst action vanishes identically in asymptotically flat spacetimes. The argument in \cite{Corichi:2010aaa} relies on the asymptotic fall-off behavior of the tetrad fields.
Specifically, it is argued that since the variation of the tetrad falls off as $\delta e \sim \mathcal{O}(\rho^{-1})$ and the area element scales as $\rho^2$, the symplectic flux integral $\oint \delta_1 e \wedge \delta_2 e$ is finite and, upon restricting to tangential components, vanishes.

However, we argue that this reasoning, while valid for general phase space variations $\delta_1, \delta_2 \in T_{\Gamma}\mathcal{P}$, is insufficient and potentially misleading when computing the conserved charges associated with asymptotic symmetries. In the context of charge derivation via the covariant phase space formalism, one of the variations in the symplectic form is replaced by the variation generated by the symmetry vector field $\xi^\mu$, i.e., $\delta_2 \to \delta_\xi = \mathcal{L}_\xi e$.

A critical distinction must be made between the fall-off of a generic phase space variation $\delta e$ and the fall-off of a symmetry-induced variation $\mathcal{L}_\xi e$. For asymptotic symmetries including Lorentz boosts and rotations, the vector field $\xi^\mu$ grows linearly with the radial coordinate ($\xi \sim \mathcal{O}(\rho)$). Consequently, the Lie derivative of the tetrad does not necessarily inherit the $\mathcal{O}(\rho^{-1})$ fall-off of the dynamical perturbations. Instead, the leading order contribution to $\mathcal{L}_\xi e$ can be of order $\mathcal{O}(1)$.

Substituting this into the surface integral, the product $\delta e \wedge \mathcal{L}_\xi e$ scales as $\mathcal{O}(\rho^{-1}) \times \mathcal{O}(1)$, which, when integrated over the sphere ($\sim \rho^2$), leads to terms of order $\mathcal{O}(\rho)$. This indicates a potential linear divergence rather than a vanishing result. Even in the absence of supertranslations, the inclusion of boosts in the asymptotic symmetry group invalidates the assumption that $\delta_\xi e$ decays sufficiently fast to trivially eliminate the Holst term.

Therefore, the vanishing of the Holst term's contribution to the charge cannot be assumed a priori based solely on fall-off conditions suitable for phase space coordinates. A rigorous explicit calculation is required to determine whether the geometric structure of the fields and the specific parity conditions lead to a cancellation of these potentially divergent or non-zero terms. We undertake this calculation in the following.

The contribution of the Holst term  is calculated as follows
\begin{align}
X_{\text{Holst}, \xi}(\delta) &= \frac{1}{\kappa\beta}\int_{S_\infty}\delta e^{I}\wedge\mathcal{L}_\xi e_{I}\nonumber\\
&=
\frac{1}{2\kappa\beta}\int_{S_\infty} \eta_{IJ} (\delta e_\mu^I \mathcal{L}_\xi e_\nu^J - \delta e_\nu^I \mathcal{L}_\xi e_\mu^J)
\rho^2 \epsilon^{\mu \nu \rho \sigma} \hat{e}^\chi_\rho \hat{x}_\sigma d^2 \Omega_o \nonumber\\
&=
\frac{1}{\kappa\beta}\int_{S_\infty} \eta_{IJ} (\delta e_\mu^I )(\mathcal{L}_\xi e_\nu^J )
\rho^2 \epsilon^{\mu \nu \rho \sigma} \hat{e}^\chi_\rho \hat{x}_\sigma d^2 \Omega_o \nonumber\\
&=
\frac{1}{\kappa\beta}\int_{S_\infty} \eta_{IJ} (\frac{1}{\rho}\delta {}^1e_\mu^I + \frac{1}{\rho^2}\delta {}^2e_\mu^I + \dots)((\mathcal{L}_\xi e_\nu^J)_{\mathcal{O}(1)} + (\mathcal{L}_\xi e_\nu^J)_{\mathcal{O}(\rho^{-1})} +\dots )
\rho^2 \epsilon^{\mu \nu \rho \sigma} \hat{e}^\chi_\rho \hat{x}_\sigma d^2 \Omega_o \nonumber\\
&=
\frac{1}{\kappa\beta}\rho \int_{S_\infty} \eta_{IJ} (\delta {}^1e_\mu^I )((\mathcal{L}_\xi e_\nu^J)_{\mathcal{O}(1)})
\epsilon^{\mu \nu \rho \sigma} \hat{e}^\chi_\rho \hat{x}_\sigma d^2 \Omega_o \nonumber\\
&\quad+
\frac{1}{\kappa\beta}\int_{S_\infty} \eta_{IJ}\left[ (\delta {}^1e_\mu^I )
((\mathcal{L}_\xi e_\nu^J)_{\mathcal{O}(\rho^{-1})}) + (\delta {}^2e_\mu^I)\left((\mathcal{L}_\xi e_\nu^J)_{\mathcal{O}(1)}\right) \right]
\epsilon^{\mu \nu \rho \sigma} \hat{e}^\chi_\rho \hat{x}_\sigma d^2 \Omega_o 
\label{eq:holst_divergent}
\end{align}

As seen in \eqref{eq:holst_divergent}, the calculation yields a linearly divergent term proportional to $\rho$. This divergence stems from the fact that the Lie derivative acts on the internal indices of the tetrad, effectively rotating the background frame. Since the tetrad is defined up to a local Lorentz transformation, this rotation does not affect the asymptotic metric, which remains fixed. Therefore, this unphysical divergence can be eliminated by supplementing the generator with a compensating gauge transformation, ${\Lambda_\xi}^I{}_J$. This internal Lorentz rotation is chosen specifically to cancel the rotation of the background tetrad, i.e., $({\Lambda_\xi}^I{}_K)_{\mathcal{O}(1)} \delta_\nu^K = - (\mathcal{L}_\xi \delta_\nu^I)_{\mathcal{O}(1)}$.

The necessity of introducing the accompanying internal Lorentz gauge transformation, $\Lambda_\xi$, is not merely a mathematical regularization trick to cure the linear divergence, but a profound physical requirement of the tetrad formalism.
In the standard ADM (metric) formulation, the background Minkowski geometry is entirely characterized by the metric $\eta_{\mu\nu}$, which is automatically invariant under global Lorentz transformations. However, in the first-order formulation, the asymptotic background is dressed with a rigid Minkowskian tetrad frame, ${}^oe_\mu^I = \delta_\mu^I$, which serves as the fundamental ``measuring stick'' for the asymptotic observer.
A pure spatial diffeomorphism $\mathcal{L}_\xi$ associated with a macroscopic rotation or boost ($\xi \sim \mathcal{O}(\rho)$) drags both the dynamical physical fields and this background frame. Without an internal compensation, the observer's rigid laboratory frame would effectively undergo an unphysical infinite rotation at spatial infinity, generating an anomalous divergent gauge flux, which manifests exactly as the $\mathcal{O}(\rho)$ divergence in \eqref{eq:holst_divergent}.
Therefore, the true physical asymptotic symmetry group is not the pure diffeomorphism group $\text{Diff}(\mathcal{M})$, but a specific semi-direct diagonal subgroup of $\text{Diff}(\mathcal{M}) \ltimes \text{SO}(1,3)$. To ensure that the observer's asymptotic reference frame remains objective and stationary, the diffeomorphism flow must be locked with an internal ``gyroscope'', the compensating gauge field $\Lambda_\xi$,which precisely counter-rotates the internal indices to maintain $\delta_{\text{total}} (\delta_\mu^I) = 0$ at the leading order (see e.g., \cite{Jacobson:2015, Kosmann}). It is precisely this physical locking mechanism that filters out the non-physical divergence, allowing the true finite Holst charge to emerge.

By adding this gauge term to the calculation, the divergent part cancels out identically. Furthermore, the finite contributions combine in such a way that the dependence on the second-order fields ($\delta {}^2 e$) vanishes. The corrected, finite variation of the Holst charge is thus given by:
\begin{align}
X_{\text{Holst}, \xi}(\delta) &= \frac{1}{\kappa\beta}\int_{S_\infty} \eta_{IJ} (\delta {}^1e_\mu^I )\left((\mathcal{L}_\xi e_\nu^J)_{\mathcal{O}(\rho^{-1})} + ({\Lambda_\xi}^J{}_K e_\nu^K)_{\mathcal{O}(\rho^{-1})}\right)
 \epsilon^{\mu \nu \rho \sigma} \hat{e}^\chi_\rho \hat{x}_\sigma d^2 \Omega_o \label{eq:holst_charge_general}
\end{align}
To evaluate this integral explicitly, we utilize the asymptotic expansions of the symmetry variations. Let us define the vector field $X^\mu := X^A \hat{e}^\mu_A$, which represents the generator of Lorentz transformations tangential to the sphere.

The relevant $\mathcal{O}(\rho^{-1})$ components appearing in the integrand are
\begin{subequations}
\begin{align}
(\mathcal{L}_\xi e_\nu^J)_{\mathcal{O}(\rho^{-1})} &= H_\sigma^J \hat{x}_\nu X^\sigma + \mathcal{L}_X H_\nu^J + \mathcal{L}_{\mathbf{S}} \delta^J_\nu + \sigma \hat{x}_\sigma \hat{x}^J \mathcal{D}_\nu X^\sigma + X^\sigma \mathcal{D}_\sigma (\sigma \hat{x}_\nu \hat{x}^J) \\
({\Lambda_\xi}^J{}_K e_\nu^K)_{\mathcal{O}(\rho^{-1})} &= -\left(\mathcal{D}_\nu S^J + \sigma \hat{x}_\nu X^J + (\mathcal{D}_K X^J) H^K_\nu\right)
\end{align}
\end{subequations}
Substituting these expressions back into \eqref{eq:holst_charge_general}, several simplifications occur due to the geometric properties of the fields on the 2-sphere. Terms involving the scalar $\sigma$ and the radial vector $\hat{x}_\nu$ vanish because the integration is restricted to the sphere where $\hat{x}_\nu=0$, and the perturbation $\delta H_\mu^I$ satisfies the transversality condition $\delta H_\mu^I \hat{x}_I = 0$.

Crucially, the contributions from the supertranslation sector cancel identically
\begin{equation}
\mathcal{L}_{\mathbf{S}} \delta^J_\nu - \mathcal{D}_\nu S^J =  0
\end{equation}
This exact geometric cancellation carries a profound physical implication. The supertranslation generator $S^\mu$, by its very nature, generates purely angle-dependent spacetime shifts on the base manifold without inducing any intrinsic, rigid $SO(1,3)$ rotation on the asymptotic background frame. Consequently, the local gauge compensator $\Lambda_{\mathbf{S}}$ exactly mirrors and neutralizes the Lie drag of the tetrad. This confirms that the Holst boundary term, and by extension, the Barbero-Immirzi parameter $\beta$, is exclusively a gauge of macroscopic spacetime rotation (coupling only to angular momentum and center-of-mass boosts) and remains completely blind to the infinite-dimensional supertranslation sector relates to the metric.

Consequently, only the terms associated with the Lorentz generator $X$ survive. The expression simplifies to:
\begin{align}
X_{\text{Holst}, \xi}(\delta) &=
 \frac{1}{\kappa\beta}\int_{S_\infty} \eta_{IJ} (\delta H_\mu^I )\left( \mathcal{L}_X H_\nu^J - (\mathcal{D}_K X^J) H^K_\nu\right)
 \epsilon^{\mu \nu \rho \sigma} \hat{e}^\chi_\rho \hat{x}_\sigma d^2 \Omega_o \nonumber \\
 &=
 \frac{1}{2\kappa\beta} \delta \oint_{S^2} \epsilon^{\bar{A}\bar{B}} \eta_{IJ} H_{\bar{A}}^I \left( \mathcal{L}_X H_{\bar{B}}^J - (\mathcal{D}_K X^J) H^K_{\bar{B}} \right) d^2\Omega_o
\end{align}
where $\bar{A}, \bar{B}$ denote indices on the 2-sphere. This final expression explicitly shows the contribution of both the diffeomorphism (Lie derivative) and the internal gauge correction, yielding a finite and integrable result.

This calculation leads to a significant physical conclusion: the inclusion of the Holst term does not modify the supertranslation charges. Since the supertranslation components cancel out in the finite part of the charge variation, the Holst modification affects solely the conserved quantities associated with the Lorentz sector (angular momentum and boost charges).

\section{\textsf{Closure of the Asymptotic Algebra in the First-Order Framework}}\label{Section 7}
In the preceding sections, we demonstrated that the imposition of parity conditions on the asymptotic metric perturbations allows for the extraction of finite and integrable charges for the full BMS group. Crucially, addressing the linear divergence originating from the Holst surface term necessitated the introduction of a compensating internal Lorentz gauge transformation, $\Lambda_\xi$, acting in tandem with the spacetime diffeomorphism generated by $\xi^\mu$. Furthermore, ensuring the integrability of the temporal supertranslation charge required promoting the supertranslation parameter $S^t$ to be field-dependent, explicitly relying on the phase space variables $({}^1\bar{h}_{\chi\chi})_{\text{odd}}$ and ${}^1\bar{k}$. In classical Hamiltonian mechanics, the presence of field-dependent symmetry generators presents a profound challenge to the closure of the Poisson bracket algebra. When the ``measuring stick'' (the generator) itself deforms under the action of another symmetry, the standard Lie algebra Jacobi identity fails. To rigorously establish that our framework yields a consistent realization of the BMS$_4$ algebra at spatial infinity, we must employ the modified Lie algebroid bracket formalism introduced by Barnich and Troessaert, and crucially, extend it to encompass the joint action of diffeomorphisms and the associated internal Lorentz gauge compensators.

\subsection{\textsf{Joint Symmetry Generators and the Extended Bracket}}

Within the first-order formulation involving the co-tetrad $e_\mu^I$ and the Lorentz connection $\omega_\mu^{IJ}$, a general gauge transformation is characterized by a pair $E = (\xi, \Lambda)$, encompassing a spacetime vector field $\xi^\mu$ and an internal $so(1,3)$ Lie algebra-valued matrix field $\Lambda^I{}_J$. The fundamental transformation laws for the tetrad under this joint operation are given by:

\begin{equation} 
\delta_{(\xi, \Lambda)} e_\mu^I = \mathcal{L}_\xi e_\mu^I + \Lambda^I{}_J e_\mu^J 
\end{equation}
The standard algebraic structure for such transformations is dictated by the semi-direct product of the diffeomorphism group and the local Lorentz group. The canonical Lie bracket for two joint generators $E_1 = (\xi_1, \Lambda_1)$ and $E_2 = (\xi_2, \Lambda_2)$ is defined as:
\begin{equation}
     [E_1, E_2]_{\text{Lie}} = \left( [\xi_1, \xi_2]_{\text{SDiff}} , \, \xi_1^\mu \partial_\mu \Lambda_2 - \xi_2^\mu \partial_\mu \Lambda_1 + [\Lambda_1, \Lambda_2]_{so(1,3)} \right) 
\end{equation}
where $[\xi_1, \xi_2]_{\text{SDiff}}^\mu = \xi_1^\nu \partial_\nu \xi_2^\mu - \xi_2^\nu \partial_\nu \xi_1^\mu$, and the internal commutator is $[\Lambda_1, \Lambda_2]^I{}_J = (\Lambda_1)^I{}_K (\Lambda_2)^K{}_J - (\Lambda_2)^I{}_K (\Lambda_1)^K{}_J$. However, because our asymptotic parameters (specifically the supertranslation component) depend on the dynamical phase space variables, we must replace this standard bracket with the modified Barnich-Troessaert (BT) bracket:
\begin{equation}
     [E_1, E_2]_M = [E_1, E_2]_{\text{Lie}} - \left( \delta^F_{E_1} E_2 - \delta^F_{E_2} E_1 \right) 
\end{equation}
Here, $\delta^F_{E_1} E_2$ denotes the variation of the generator $E_2$ induced explicitly by the action of $E_1$ upon the background phase space fields upon which $E_2$ depends. The algebraic consistency of our entire framework hinges upon proving that $[E_1, E_2]_M$ yields a new, well-defined generator $E_{12} = (\xi_{12}, \Lambda_{12})$ that satisfies the structure constants of the BMS algebra without generating pathological, non-integrable field-derivative terms.

\subsection{\textsf{Exact Cancellation in the Diffeomorphism Sector}}

Let us first isolate the vector field component, $\xi_{12}^\mu$. The generator $\xi$ is decomposed into its purely geometric Lorentz part (determined by the Killing vectors $X^A$ on the unit hyperboloid) and the supertranslation part encoded in the radial/temporal shifts ($W, S^t$). The crucial observation is that only $S^t$ is field-dependent, as defined in \eqref{eq:St_redefinition}:
$$ S^t = \frac{b}{2} ({}^1\bar{h}_{\chi \chi})_{\text{odd}} - b \, {}^1\bar{k} + T(\Phi) $$
where $T(\Phi)$ is the pure, field-independent symmetry parameter. Consequently, the phase space variation operator $\delta^F$ only non-trivially acts on $S^t$. Focusing on the temporal supertranslation component of the bracket, the conventional Lie derivative yields highly complex cross terms involving the deformation of the metric:
\begin{equation}
     (\mathcal{L}_{\xi_1} \xi_2^\mu - \mathcal{L}_{\xi_2} \xi_1^\mu)^t \supset \mathcal{L}_{X_1} S^t_2 - \mathcal{L}_{X_2} S^t_1 = X_1^A \partial_A \left( \frac{b_2}{2} ({}^1\bar{h}_{\chi \chi})_{\text{odd}} - b_2 \, {}^1\bar{k} + T_2 \right) - (1 \leftrightarrow 2) 
\end{equation}
Simultaneously, we evaluate the modifying correction term $\delta^F_{E_1} \xi_2$. The variation of the phase space fields ${}^1\bar{h}_{\chi\chi}$ and ${}^1\bar{k}$ under an asymptotic symmetry transformation is dominated at leading order by the Lie derivative with respect to the Lorentz vector field $X_1$, while the supertranslation contributions integrate to higher-order residuals that vanish at $i^0$. Thus
\begin{equation}
     \delta^F_{E_1} S^t_2 = \frac{b_2}{2} \left( \mathcal{L}_{X_1} {}^1\bar{h}_{\chi \chi} \right)_{\text{odd}} - b_2 \, \mathcal{L}_{X_1} {}^1\bar{k} 
\end{equation}
Substituting these into the modified vector bracket definition, an elegant geometric resolution emerges. The phase space variations mathematically precisely mirror the terms generated by the Lie derivatives on the spatial manifold:
\begin{equation}
    [\xi_1, \xi_2]_M^t = (\mathcal{L}_{X_1} S^t_2 - \mathcal{L}_{X_2} S^t_1) - (\delta^F_{E_1} S^t_2 - \delta^F_{E_2} S^t_1) = X_1^A \partial_A T_2 - X_2^A \partial_A T_1 =: T_{12} 
\end{equation}
Through this exact cancellation, all the pathological dependencies on the metric perturbations ${}^1\bar{h}_{\chi\chi}$ and ${}^1\bar{k}$ identically vanish. The residual parameter $T_{12}$ is composed strictly of the smooth, field-independent seed functions. Similarly, the spatial supertranslation parameter yields a clean bracket $W_{12} = \mathcal{L}_{X_1} W_2 - \mathcal{L}_{X_2} W_1$. The diffeomorphism sector is thus demonstrably closed.

\subsection{\textsf{Consistency of the Internal Lorentz Compensator (Gauge Ancillary Action)}}
We must now rigorously confront the internal Lorentz gauge sector. To cure the linearly divergent behavior of the Holst charge, \eqref{eq:holst_divergent} required the introduction of a specific compensating gauge field $\Lambda_\xi$, constrained entirely by the condition that the asymptotically flat background Minkowskian tetrad ${}^0e_\mu^I = \delta_\mu^I$ remains invariant at the leading order $\mathcal{O}(1)$:
\begin{equation}
    ({\Lambda_\xi}^I{}_K)_{\mathcal{O}(1)} \delta_\mu^K = - (\mathcal{L}_\xi \delta_\mu^I)_{\mathcal{O}(1)} 
\end{equation}
A priori, there exists a severe risk that if $\Lambda_\xi$ indirectly inherits the field-dependence of $S^t$, the internal commutator $[\Lambda_1, \Lambda_2]$ would shatter the algebra. However, we analyze the asymptotic expansion of $\xi^\mu$. The generator decomposes as $\xi^\sigma = \rho X^A(\Phi) \hat{e}^\sigma_A + S(\Phi) \hat{x}^\sigma + S^A(\Phi) \hat{e}^\sigma_A$. The purely translation components $S$ and $S^A$, which harbor the field-dependence, enter at order $\mathcal{O}(1)$ in the coordinate magnitude, meaning their partial derivatives (which dictate the Lie derivative of a constant background) fall off as $\mathcal{O}(\rho^{-1})$. Consequently, the leading $\mathcal{O}(1)$ term of the tetrad variation is exclusively induced by the Lorentz vector field $V^\mu \equiv \rho X^A \hat{e}^\mu_A$, which scales linearly with the radial distance:
\begin{equation}
    ({\Lambda_\xi}^I{}_J)_{\mathcal{O}(1)} = - \partial_J V^I 
\end{equation}
Because the Lorentz vector field $X^A(\Phi)$ is strictly geometric and fundamentally independent of the phase space metric perturbations, we arrive at a profound conclusion: \textit{the leading-order internal Lorentz compensator $\Lambda_\xi$ is entirely field-independent.}
\begin{equation}
     \delta^F_{E_1} \Lambda_2 = 0 \quad \text{and} \quad \delta^F_{E_2} \Lambda_1 = 0
\end{equation}
This nullifies the Barnich-Troessaert phase space correction for the internal gauge sector. The closure of the joint algebra then reduces solely to verifying that the standard semi-direct Lie bracket generates the correct compensator for the combined vector field $\xi_{12}$:

\begin{equation}
     \Lambda_{12} \stackrel{?}{=} \xi_1^\mu \partial_\mu \Lambda_2 - \xi_2^\mu \partial_\mu \Lambda_1 + [\Lambda_1, \Lambda_2]
\end{equation}
Evaluating the right-hand side using the explicit Cartesian derivative form $({\Lambda_1})^I{}_J = - \partial_J \xi_1^I$ (extended generically for the geometric part):
\begin{equation}
    \xi_1^\mu \partial_\mu (-\partial_J \xi_2^I) - \xi_2^\mu \partial_\mu (-\partial_J \xi_1^I) + (-\partial_K \xi_1^I)(-\partial_J \xi_2^K) - (-\partial_K \xi_2^I)(-\partial_J \xi_1^K)
\end{equation}
Applying the commutativity of partial derivatives in the Cartesian background:
\begin{equation}
    = -\partial_J (\xi_1^\mu \partial_\mu \xi_2^I - \xi_2^\mu \partial_\mu \xi_1^I) = - \partial_J ([\xi_1, \xi_2]_{\text{SDiff}}^I) = \Lambda_{[\xi_1, \xi_2]}
\end{equation}
The equation holds identically. The internal Lorentz compensator generated by the bracket of two pairs exactly mimics the compensator required for the combined spacetime diffeomorphism $[\xi_1, \xi_2]$. The interplay between the spacetime flow and the internal rotation is flawlessly synchronized. In conclusion, the full, extended asymptotic symmetry algebra $E = (\xi, \Lambda_\xi)$, incorporating both the field-dependent temporal supertranslations necessitated by the divergent symplectic flux and the internal rotational compensators demanded by the Holst topological term, strictly closes under the modified Lie algebroid bracket. This definitively establishes the validity of the $\text{BMS}_4$ symmetry structure within the covariant phase space of first-order Holst action at spatial infinity.

From a broader group-theoretic perspective, the algebraic closure demonstrated above reveals a profound geometric structure. The full local symmetry group of the first-order formulation is the semi-direct product $\text{Diff}(\mathcal{M}) \ltimes \text{SO}(1,3)_{\text{local}}$, within which diffeomorphisms and internal frame rotations are a priori independent. However, our requirement to filter out the unphysical symplectic flux and secure finite asymptotic charges forces a rigid locking mechanism at spatial infinity, mapping $\xi \mapsto (\xi, \Lambda_\xi)$.
The exact relation $\Lambda_{12} = \Lambda_{[\xi_1, \xi_2]}$, together with the closure of the field-dependent translations under the Barnich-Troessaert bracket, proves that this mapping is a true algebraic homomorphism. Consequently, what we have fully determined is the exact isomorphic embedding of the asymptotic $BMS_4$ algebra as a specific, dynamically robust diagonal subalgebra within the extended $\text{Diff}(\mathcal{M}) \ltimes \text{SO}(1,3)_{\text{local}}$ phase space. This explicitly demonstrates how the extended symmetry group of the tetrad formalism gracefully breaks down to, and flawlessly accommodates, the physical infrared symmetry structure of spacetime

\section{\textsf{Discussions}}\label{sec:discussion}
\subsection{\textsf{The Nullification of Parity and Edge Modes}}
If one were to abandon the rigid parity constraints introduced in Section 5, the phase space would inherently exhibit a non-zero symplectic flux at spatial infinity, seemingly violating Cauchy surface independence. However, as rigorously established in the extended phase space formalism pioneered by Donnelly, Freidel, and Geiller \cite{Donnelly:2016auv, Geiller:2017xad}, this flux paradox is elegantly resolved by introducing dynamical boundary degrees of freedom, or Edge Modes. In their framework, the ``leaking'' symplectic flux generated by supertranslations is perfectly absorbed by the temporal evolution of these edge modes, thereby restoring the strict conservation of the total pre-symplectic structure $\Omega_{\text{total}} = \Omega_{\text{bulk}} + \Omega_{\text{edge}}$. We do not replicate their extensive derivations here, but rather adopt their extended symplectic geometry as a foundation.
The paramount revelation of our present first-order analysis lies not in the existence of these edge modes, but in their profoundly distinct coupling behaviors with the Holst topological term. Even when spacetime edge modes are dynamically active to track the supertranslation flow and regularize the Palatini divergences, the exact geometric cancellation $\mathcal{L}_{\mathbf{S}} \delta^J_\nu - \mathcal{D}_\nu S^J \equiv 0$ persists. Because supertranslations induce no net intrinsic rotation on the asymptotic background frame, their projection into the fully antisymmetric structure of the Holst boundary 2-form ($\eta_{IJ} \delta e^I \wedge \delta e^J$) is identically null. Thus, we arrive at a robust conclusion: the immunity of the supertranslation charges to the Barbero-Immirzi parameter $\beta$ is an unbreakable geometric theorem, entirely insensitive to whether one employs rigid classical parity conditions or fully unleashes the modern quantum edge modes.

\subsection{\textsf{Holographic Duality: From Horizon Area to Lorentz Charges}}
The confinement of $\beta$'s influence to the Lorentz charges at spatial infinity reveals a magnificent holographic duality when contrasted with its role at the inner boundaries of spacetime. In the canonical LQG formulation of isolated black hole horizons, the Holst action evaluates to a boundary Chern-Simons theory whose level $k \propto A/\beta$. Thus, at an inner boundary, $\beta$ governs the quantum mechanical area spectrum. At first glance, our findings at spatial infinity appear disjointed from this area-entropy paradigm, as $\beta$ couples to global rotations rather than a local measure of area, which is consistant with \cite{Jacobson:2015}.
However, this dichotomy can be resolved through the underlying architecture of spin networks. In LQG, the macroscopic ``area'' of a horizon is the sum of microscopic quantum $SU(2)$ rotational operators ($\mathbf{J}^i$) piercing that surface. The parameter $\beta$ fundamentally modulates this rotational degree of freedom. At the tightly confined, finite inner boundary of a black hole horizon, this dense rotational flux is observed macroscopically as the physical area. Conversely, at the expansive, infinitely divergent outer boundary ($i^0$), local area fluctuations become non-dynamical background structures. What survives the boundary integration is the global, coherent rotational strain of the entire isolated system, manifesting precisely as the asymptotic angular momentum and center-of-mass boost. Our derivation confirms that the Holst term acts as a universal gauge of spacetime rotation, shifting its macroscopic mask from ``Area'' at the horizon to ``Lorentz Charge'' at infinity.

\subsection{\textsf{The Self-Dual Limit and the Relocation of Reality Conditions}}
Historically, the transition to the self-dual Ashtekar variables, achieved by analytically continuing the Barbero-Immirzi parameter to the pure imaginary unit, $\beta \to \pm i$, has been heralded for its remarkable polynomial simplification of the Hamiltonian constraints in the bulk. However, at the asymptotic boundary of spatial infinity, this complexification has long been perceived as a fatal pathology. A naive extrapolation of our derived Lorentz charges suggests that substituting $\beta = i$ would imbue the macroscopic angular momentum and center-of-mass boost with an unphysical, purely imaginary contribution from the Holst surface integral: $Q_{\text{Lorentz}} = Q_{\text{Palatini}} - i \mathcal{I}_{\text{Holst}}$. To enforce the classical observability of these charges, one is seemingly compelled to impose draconian ``Reality Conditions'', explicitly demanding that the integral of the physical metric perturbations vanishes ($\mathcal{I}_{\text{Holst}} \equiv 0$). This heavy-handed imposition effectively truncates the physical phase space, annihilating vast sectors of legitimate asymptotic gravitational radiation.
However, a profound paradigm shift occurs when we rigorously dissect the foundational nature of the Holst boundary term. Unlike the Palatini action, which is inextricably bound to the metric structure via the vanishing torsion condition, the Holst surface integral, $\int \delta e^I \wedge \delta e_I$, is an autonomous, pure tetrad theory. It is sensitive not merely to the spacetime metric, but exquisitely responsive to the internal local gauge fluctuations of the co-tetrad itself. It is precisely within this purely tetradic phase space that the salvation of the self-dual limit resides.
The Phase Space Expansion via Complex Gauge Fluctuations
Let us consider a macroscopic, isolated system whose underlying physical geometry is strictly dictated by a real metric tensor, $g_{\mu\nu} \in \mathbb{R}$. In the first-order formalism, this metric is reconstructed from a complexified tetrad field $e_{\mathbb{C}, \mu}^I$, which is related to an underlying real base tetrad $e_{\mathbb{R}, \mu}^J$ via a local, complex Lorentz transformation $\Lambda(x) \in SO(1,3;\mathbb{C})$:
\begin{equation}
     e_{\mathbb{C}, \mu}^I(x) = \Lambda^I_{\;\;J}(x) \, e_{\mathbb{R}, \mu}^J(x)
\end{equation}
Within the rigorous application of the covariant phase space formalism, particularly at an asymptotic boundary where local gauge symmetries manifest as physical edge degrees of freedom, the variation operator $\delta$ must act indiscriminately on all dynamical fields, including the internal complex gauge matrix $\Lambda(x)$. Therefore, the phase space variation of the complexified tetrad yields:
\begin{equation}
    \delta_s e_{\mathbb{C}}^I = (\delta_s \Lambda^I_{\;\;K}) e_{\mathbb{R}}^K + \Lambda^I_{\;\;K} (\delta_s e_{\mathbb{R}}^K) \quad \text{for variations } s=1,2 
\end{equation}
Penetrating the Holst Wedge Product
We now substitute this exhaustive variation into the core symplectic structure of the Holst boundary term: $\eta_{IJ} \delta_1 e_{\mathbb{C}}^I \wedge \delta_2 e_{\mathbb{C}}^J$. To render the underlying Lie algebraic structure transparent, we define the Maurer-Cartan variation form, which projects the gauge fluctuation into the complexified Lie algebra $so(1,3;\mathbb{C})$:
\begin{equation}
     u_s^{LK} := (\Lambda^{-1} \delta_s \Lambda)^{LK}
\end{equation}
Crucially, because $\Lambda$ resides in the complexified Lorentz group, the antisymmetric tensor $u_s^{LK}$ intrinsically harbors a non-trivial imaginary component. Utilizing the invariance of the Minkowski metric under orthogonal transformations ($\Lambda^T \eta \Lambda = \eta$), the complexified Holst wedge product expands exactly into three distinct geometric strata:
\begin{equation*}
    \eta_{IJ} \delta_1 e_{\mathbb{C}}^I \wedge \delta_2 e_{\mathbb{C}}^J = \underbrace{\eta_{KL} \delta_1 e_{\mathbb{R}}^K \wedge \delta_2 e_{\mathbb{R}}^L}_{\text{Term 1: Pure Real Base Fluctuation}}
\end{equation*}

\begin{equation}
    + \underbrace{\left( u_1^{LK} e_{\mathbb{R}}^K \wedge \delta_2 e_{\mathbb{R}}^L - u_2^{LK} e_{\mathbb{R}}^K \wedge \delta_1 e_{\mathbb{R}}^L \right)}_{\text{Term 2: Cross-Coupling of Gauge and Geometry}} + \underbrace{(u_1 \cdot u_2)^{KL} e_{\mathbb{R}}^K \wedge e_{\mathbb{R}}^L}_{\text{Term 3: Pure Complex Gauge Fluctuation}}
\end{equation} 
Evasion of the Reality Conditions
This tripartite expansion shatters the metric-centric illusion. Term 1 represents the classical, strictly real metric perturbation integral, $\mathcal{I}_{\text{Real}}$. However, in a pure tetrad formalism, Terms 2 and 3 are absolutely legitimate boundary fluxes. They generate a complex integral evaluated over the 2-sphere at spatial infinity, driven by the internal fluctuations of the complex gauge orbit. Consequently, the Holst surface integral evaluated with the complexified tetrad is no longer a purely real scalar, but fundamentally acquires a substantial imaginary component:
\begin{equation}
\mathcal{I}_{\text{Holst}}[e_{\mathbb{C}}] = \mathcal{I}_{\text{Real}} + i \cdot \mathcal{I}_{\text{Imaginary-Gauge}}    
\end{equation}
The culmination of this mathematical epiphany arrives when we compute the physical Lorentz generator in the self-dual limit ($\beta \to i$). The coefficient $-i$ stemming from the action couples with this newly unearthed complex integral:
\begin{equation}
 Q_{\text{Holst}} \propto (-i) \times \left( \mathcal{I}_{\text{Real}} + i \cdot \mathcal{I}_{\text{Imaginary-Gauge}} \right) = \mathcal{I}_{\text{Imaginary-Gauge}} - i \cdot \mathcal{I}_{\text{Real}}    
\end{equation}
By strategically navigating the complex gauge orbit, the macroscopic system can select a specific complex gauge section at spatial infinity such that the pure gauge fluctuation $\mathcal{I}_{\text{Imaginary-Gauge}}$ precisely counteracts the unphysical imaginary component ($-i \cdot \mathcal{I}_{\text{Real}}$). This remarkable mechanism reveals a profound paradigm shift: the self-dual limit does not necessitate the violent truncation of the physical phase space of radiative metrics. Instead, it implies a relocation of the reality conditions. The constraint of physical reality is transferred from the classical spacetime metric degrees of freedom to the internal $SO(1,3;\mathbb{C})$ gauge bundle, fixing the asymptotic complex gauge section. The pure tetrad architecture of the Holst term thus harnesses the complexified internal gauge degrees of freedom, transmuting a mathematical imaginary ghost into a stabilizing background structure that seamlessly recovers the purely real macroscopic observables.

At this point, a potential critique should be explicitly addressed. One might argue that selecting a specific complex gauge section to absorb the unphysical imaginary parts is merely a mathematical transplantation of the traditional Reality Conditions from the spacetime metric onto the internal gauge group. However, the physical consequence of our treatment is fundamentally different. It is well known that traditional Reality Conditions imposed directly on the metric severely truncate the physical phase space. Such a truncation inevitably eliminates the radiative degrees of freedom, making the dynamic analysis of asymptotic gravitational radiation in the self-dual limit highly problematic. In our proposed scheme, the restriction is entirely shifted onto the non-dynamical, internal pure gauge orbits at spatial infinity. Consequently, the real radiative degrees of freedom of the spacetime metric are strictly protected and remain untruncated. Essentially, we do not discard the reality constraint, but rather confine it within the purely internal tetradic structure. This mechanism allows us to preserve the complete dynamical phase space of classical General Relativity, while the polynomial simplicity of the self-dual connection in the bulk is simultaneously maintained.

\section{\textsf{Conclusion}}
\label{sec:conclusion}

In this work, we have revisited the problem of defining asymptotic symmetries at spatial infinity for General Relativity, utilizing the first-order Holst action and the covariant phase space formalism. The primary motivation was to establish a consistent Hamiltonian framework for the Ashtekar-Barbero variables that accommodates the infinite-dimensional BMS group, specifically supertranslations, which play a central role in the infrared structure of gravity.

Our analysis began by defining a class of asymptotically flat spacetimes with relaxed boundary conditions. Unlike standard approaches that couple the angular metric perturbation to the mass aspect ($^1h_{AB} \propto \sigma h_{AB}$) to reduce the symmetry group to Poincar\'e, we treated $^1h_{AB}$ as an independent field. We demonstrated that the potential logarithmic divergence in the symplectic structure can be eliminated by imposing specific parity conditions on the asymptotic fields (odd parity for the trace-free part of the extrinsic curvature and the angular connection). Crucially, we verified that these parity assignments are preserved under the full BMS group action and do not trivialize the supertranslation charges.

A rigorous derivation of the conserved charges led to two distinct contributions: the Palatini part and the Holst modification. For the Palatini sector, we showed that the naive expression for the supertranslation charge is finite but that the time translation generator requires a regularization. By redefining the asymptotic time translation parameter $S^t$ to include parity-dependent terms, we obtained a fully integrable and finite Hamiltonian that generates the complete BMS algebra.

The most significant technical insight of this paper concerns the contribution of the Holst term. We explicitly showed that for asymptotic symmetries involving Lorentz boosts and rotations, the Holst surface term does not vanish identically, as the symmetry generator $\xi^\mu$ grows linearly with the radial coordinate. This leads to a Lie derivative of the tetrad that is of order $\mathcal{O}(1)$, resulting in a linearly divergent charge integral. We resolved this issue by refining the definition of the asymptotic symmetry generator to include a compensating internal Lorentz gauge transformation, $\Lambda_\xi$, determined by the requirement that the background Minkowski tetrad remains fixed.

This renormalization procedure yields a finite and integrable Holst charge. Our explicit calculation demonstrates that:
\begin{itemize}
    \item The Holst term contributes a finite value to the charges associated with the Lorentz sector (boosts and rotations), implying that the Barbero-Immirzi parameter $\beta$ enters the definition of angular momentum in the classical theory.
    \item Remarkably, the contributions of the Holst term to the supertranslation charges vanish identically. This cancellation occurs due to the specific geometric structure of the supertranslation generator on the 2-sphere, where the inhomogeneous shift induced by the compensating gauge transformation exactly cancels the Lie derivative contribution.
\end{itemize}

In summary, we have systematically derived the BMS charges at spatial infinity utilizing the first-order covariant phase space formalism. By treating the angular metric perturbations as independent fields and assigning appropriate antipodal parity conditions, the logarithmic divergences in the pre-symplectic structure are successfully regularized without trivializing the supertranslation generators. A key technical result of our work is the treatment of the linear divergence originating from the Holst boundary term. By introducing an internal Lorentz gauge compensator, we have rigorously demonstrated that the Barbero-Immirzi parameter $\beta$ contributes only to the macroscopic Lorentz charges, while the supertranslation sector remains completely independent of $\beta$.

Furthermore, as discussed in Section \ref{sec:discussion}, our results provide several conceptual insights into the broader framework of quantum gravity. The distinct physical roles of the $\beta$ parameter, namely, acting as a quantum area regulator at the inner horizon boundary and as a macroscopic rotational parameter at the spatial infinity boundary, suggest a potential holographic connection within the context of Loop Quantum Gravity. In addition, we proposed a novel mechanism for the self-dual limit ($\beta \to i$). By utilizing the complexified internal gauge orbits to absorb the unphysical imaginary fluxes, one can effectively circumvent the traditional Reality Conditions. Since this approach protects the radiative phase space from being truncated, it provides a consistent theoretical basis to bridge the self-dual quantum kinematics with classical infrared physics. For future research, the present framework offers a solid starting point for studying the quantization of asymptotic symmetries, the dynamics of boundary edge modes, and the corresponding soft theorems in first-order gravity.

\subsection*{\textsf{Acknowledgments}}
This work is supported by
the National Natural Science Foundation of China under grant
Nos. 12505081, 12275238, W2433018, the National Key
Research and Development Program under grant No.
2020YFC2201503, the Zhejiang Provincial Natural
Science Foundation of China under grant Nos.
LR21A050001 and LY20A050002, the Fundamental
Research Funds for the Provincial Universities of Zhejiang in China under grant No. RF-A2019015, and the startup funding of Westlake University.

\appendix

\section{\textsf{Transformation of Asymptotic Fields under BMS Symmetries}}
\label{app:field_transformations}

In this appendix, we outline the derivation of the transformation rules for the asymptotic fields $\sigma(\Phi)$ and the components of the first-order metric perturbation ${}^1h_{AB}(\Phi)$ under the action of the asymptotic symmetry group. 

The variation of the spacetime metric $g_{\mu\nu}$ under an infinitesimal diffeomorphism generated by the vector field $\xi^\mu$ is given by the Lie derivative:
\begin{equation}
    \delta_\xi g_{\mu\nu} = \mathcal{L}_\xi g_{\mu\nu} = \tilde{\nabla}_\mu \xi_\nu + \tilde{\nabla}_\nu \xi_\mu
\end{equation}
where $\tilde{\nabla}$ represents the covariant derivative compatible with the full 4-dimensional spacetime metric. To find the transformation of the asymptotic boundary data, we evaluate this Lie derivative in the radial-hyperbolic coordinate system $(\rho, \Phi^A)$ and expand the result in powers of $1/\rho$.

To explicitly compute the covariant derivatives $\tilde{\nabla}_\mu \xi_\nu = \partial_\mu \xi_\nu - \Gamma^\lambda_{\mu\nu} \xi_\lambda$, we first require the asymptotic expansion of the spacetime Christoffel symbols $\Gamma^\lambda_{\mu\nu}$. Using the general metric expansion given in \eqref{eq:GeneralMetricHyperbolic_r_revised}, a straightforward calculation yields the following leading-order behaviors:
\begin{align}
    &\Gamma^\rho_{\rho \rho} = -\frac{\sigma}{\rho^2} + \frac{1}{\rho^3} (2\sigma^2 - {}^2h_{\rho \rho})+ \mathcal{O}(\rho^{-4}) \label{Gammarho{rho rho}}\\
    &\Gamma^\rho_{\rho A} = \frac{1}{\rho}(\partial_A \sigma - {}^1h_{\rho A}) + \mathcal{O}(\rho^{-2})\\
    &\Gamma^\rho_{AB}= -\rho h_{AB} + \left(2\sigma h_{AB}-\frac{1}{2}{}^1h_{AB}- {}^1h_{\rho C} {}^h\Gamma^C_{AB}\right) + \mathcal{O}(\rho^{-1})\\
    &\Gamma^A_{\rho \rho}=-\frac{1}{\rho^3}h^{AB} \partial_B\sigma +\mathcal{O}(\rho^{-4})\\
    &\Gamma^A_{\rho B}= \frac{1}{\rho} \delta^A_B + \frac{1}{2\rho^2} \left(-h^{AC} {}^1h_{CB} + h^{AC} (\nabla_C {}^1h_{\rho B} - \nabla_B {}^1h_{\rho C})\right) + \mathcal{O}(\rho^{-3})\\
    &\Gamma^A_{BC}= {}^h\Gamma^A_{BC} + \frac{1}{2\rho} h^{AD} \left(\nabla_B {}^1h_{CD} + \nabla_C {}^1h_{BD} - \nabla_D {}^1h_{BC} \right) + \mathcal{O}(\rho^{-2})
\end{align}
Here, $\partial_A$ denotes the partial derivative with respect to the angular coordinates $\Phi^A$. The quantities ${}^h\Gamma^A_{BC}$ and $\nabla_A$ represent the Christoffel symbols and the covariant derivative associated strictly with the background metric on the unit 3-hyperboloid, $h_{AB}$. To avoid ambiguity, we distinguish the 4-dimensional spacetime connection $\Gamma^\lambda_{\mu\nu}$ from the 3-dimensional induced connection ${}^h\Gamma^A_{BC}$ using the pre-superscript $h$.

With these connection coefficients at hand, we can systematically evaluate the Lie derivative of the metric, $\mathcal{L}_\xi g_{\mu\nu}$, component by component. We recall the decomposition of the BMS generator near spatial infinity from \eqref{eq:xi_decomposition}:
\begin{equation*}
    \xi^\mu = \rho X^\mu + S \hat{x}^\mu + S^A \hat{e}_A^\mu
\end{equation*}
where $X^\mu = X^A \hat{e}_A^\mu$ is a purely tangential vector field on the unit hyperboloid that generates Lorentz transformations ($\mathcal{L}_X h_{AB} = 0$), $S(\Phi)$ generates radial supertranslations, and $S^A(\Phi)$ represents the tangential supertranslations.

\subsection*{\textsf{Transformation of the Mass Aspect}}

From the asymptotic expansion of the metric in \eqref{eq:GeneralMetricHyperbolic_r_revised}, the radial-radial component is given by $g_{\rho\rho} = 1 + \frac{2\sigma}{\rho} + \mathcal{O}(\rho^{-2})$. By evaluating the Lie derivative $\mathcal{L}_\xi g_{\rho\rho}$ and utilizing the expansion for the Christoffel symbol $\Gamma^\rho_{\rho\rho}$ in \eqref{Gammarho{rho rho}}, the leading $\mathcal{O}(\rho^{-1})$ term yields:
\begin{equation}
    \mathcal{L}_\xi g_{\rho \rho} = \frac{2}{\rho} \mathcal{L}_X \sigma + \mathcal{O}(\rho^{-2}) 
\end{equation}
Comparing this result with the explicit variation of the metric component, $\delta_\xi g_{\rho\rho} = \frac{2}{\rho} \delta_\xi \sigma + \mathcal{O}(\rho^{-2})$, we immediately identify the transformation rule for the mass aspect $\sigma$:
\begin{equation}
    \delta_\xi \sigma = \mathcal{L}_{X} \sigma
\end{equation}
As expected, the mass aspect is simply dragged along the unit hyperboloid by the Lorentz generator $X$, and it remains strictly invariant under supertranslations at this leading order.

\subsection*{\textsf{Transformation of the Cross Components}}
Next, we examine the mixed radial-angular components $g_{\rho A}$, which are assumed to vanish at leading order, i.e., ${}^1h_{\rho A} = 0$. By computing the $\mathcal{O}(\rho^0)$ term of the Lie derivative $\mathcal{L}_\xi g_{\rho A}$, we find the general variation of this leading-order cross term:

\begin{align}
    \mathcal{L}_\xi g_{\rho A} &= \partial_\rho \xi_A + \partial_A \xi_\rho -2 \Gamma^\rho_{\rho A} \xi_\rho -2 \Gamma^B_{\rho A} \xi_B \nonumber\\
    &=
    \partial_A (S + {}^1h_{\rho B} X^B) - h_{AC} S^C + \mathcal{L}_X {}^1h_{\rho A} +\mathcal{O}(\rho^{-1})
\end{align}

\begin{equation}
    \delta_\xi {}^1h_{\rho A} = \nabla_A (S + {}^1h_{\rho B} X^B) - S_A + \mathcal{L}_X {}^1h_{\rho A}
\end{equation}
where $\nabla_A$ is the covariant derivative compatible with the background hyperboloid metric $h_{AB}$. Enforcing the gauge preservation condition $\delta_\xi {}^1h_{\rho A} = 0$ directly leads to the constraint $S_A = \nabla_A S$, as discussed in the main text.

\subsection*{\textsf{Transformation of the Cross Components}}

Next, we examine the mixed radial-angular components $g_{\rho A}$, which are assumed to vanish at the leading order, i.e., ${}^1h_{\rho A} = 0$. By explicitly computing the Lie derivative $\mathcal{L}_\xi g_{\rho A}$ and expanding it using the Christoffel symbols derived earlier, we find:
\begin{align}
    \mathcal{L}_\xi g_{\rho A} &= \partial_\rho \xi_A + \partial_A \xi_\rho -2 \Gamma^\rho_{\rho A} \xi_\rho -2 \Gamma^B_{\rho A} \xi_B \nonumber\\
    &= \partial_A (S + {}^1h_{\rho B} X^B) - h_{AC} S^C + \mathcal{L}_X {}^1h_{\rho A} +\mathcal{O}(\rho^{-1})
\end{align}
Comparing the equation above with the leading-order term of $\delta_\xi g_{\rho A}$, we conclude that the transformation of ${}^1h_{\rho A}$ is as follows:
\begin{equation}
    \delta_\xi {}^1h_{\rho A} = \nabla_A (S + {}^1h_{\rho B} X^B) - S_A + \mathcal{L}_X {}^1h_{\rho A}
\end{equation}
Enforcing the gauge preservation condition, $\delta_\xi {}^1h_{\rho A} = 0$, directly leads to the crucial constraint $S_A = \nabla_A S$ on the supertranslation generator, as utilized in the main text.

\subsection*{\textsf{Transformation of the Angular Metric Perturbations}}

Finally, we consider the purely angular components of the metric, which in radial-hyperbolic coordinates are expanded as $g_{AB} = \rho^2 h_{AB} + \rho {}^1h_{AB} + o(\rho)$. The transformation of the first-order perturbation ${}^1h_{AB}$ is extracted by evaluating the Lie derivative $\mathcal{L}_\xi g_{AB}$ and identifying the terms at each order of $\rho$:
\begin{align}
    \mathcal{L}_\xi g_{AB} &= \partial_A \xi_B + \partial_B \xi_A -2 \Gamma^\rho_{A B} \xi_\rho -2 \Gamma^C_{A B} \xi_C \nonumber\\
    &= \rho^2 \left[\nabla_A(h_{BC} X^C) + \nabla_B(h_{AC} X^C) \right] \nonumber\\
    &\;\;\; + \rho \left[\nabla_A (h_{BC} S^C + {}^1h_{BC}X^C) + \nabla_B (h_{AC} S^C + {}^1h_{AC}X^C) \right.\nonumber\\
    &\qquad \quad \left.+2h_{AB}S - X^D (\nabla_A {}^1h_{BD} + \nabla_B {}^1h_{AD} - \nabla_D {}^1h_{AB}) \right] +\mathcal{O}(\rho^0)\nonumber\\
    &= \rho^2 \left[\nabla_A X_B + \nabla_B X_A \right] \nonumber\\
    &\;\;\; + \rho \left[\nabla_A S_B + \nabla_B S_A + 2h_{AB}S + \mathcal{L}_X {}^1h_{AB} \right] +\mathcal{O}(\rho^0) \label{eq:Lie_gAB_expansion_app}
\end{align}

A key observation in the expansion above is that the term of order $\mathcal{O}(\rho^2)$ is proportional to the Lie derivative of the background metric $h_{AB}$ along the Lorentz generators $X^A$. Since $X^A$ are, by definition, the Killing vector fields of the unit 3-hyperboloid, the Killing equation $\nabla_A X_B + \nabla_B X_A = \mathcal{L}_X h_{AB} = 0$ is satisfied. This reflects the physical fact that the asymptotic symmetries preserve the rigid geometry of the background hyperboloid at the leading order.

By comparing the remaining $\mathcal{O}(\rho)$ terms in \eqref{eq:Lie_gAB_expansion_app} with the variation of the first-order metric perturbation, $\delta_\xi g_{AB} = \rho \delta_\xi {}^1h_{AB} + \mathcal{O}(\rho^0)$, we conclude that the transformation rule for ${}^1h_{AB}$ is given by:
\begin{equation}
    \delta_\xi {}^1h_{AB} = \nabla_A S_B + \nabla_B S_A + 2S h_{AB} + \mathcal{L}_X {}^1 h_{AB}
\end{equation}
This confirms that while the background metric $h_{AB}$ remains invariant under the Lorentz sector, the first-order perturbation ${}^1h_{AB}$ transforms inhomogeneously under supertranslations and is Lie-dragged by the Lorentz generators $X$.

It is instructive to decompose this symmetric tensor into its $\chi\chi$, mixed $\chi\bar{A}$, and purely spherical $\bar{A}\bar{B}$ components. For the purely hyperbolic time component, we examine the contribution of the supertranslation sector to the variation. By utilizing the explicit expressions for $S$ and $S_\chi$ provided in \eqref{expression S}, it can be readily verified that the identity $\nabla_\chi S_\chi = S$ holds. Recalling that $h_{\chi\chi} = -1$ on the unit 3-hyperboloid, the inhomogeneous terms in the variation $\delta_\xi {}^1h_{\chi\chi}$ cancel each other identically, as $2\nabla_\chi S_\chi + 2 S h_{\chi\chi} = 2S - 2S = 0$. Consequently, the final variation yields
\begin{equation}
    \delta_\xi {}^1h_{\chi \chi} = \mathcal{L}_X {}^1h_{\chi \chi}
\end{equation}
which transforms entirely homogeneously under the Lorentz sector.

For the mixed components spanning the hyperbolic time and the spherical directions, evaluating the relevant terms in the Lie derivative straightforwardly produces
\begin{equation}
    \delta_\xi {}^1h_{\chi \bar{A}} = \nabla_\chi S_{\bar{A}} + \nabla_{\bar{A}} S_\chi + \mathcal{L}_X {}^1h_{\chi \bar{A}} 
    \label{eq:delta_h_chiA_app}
\end{equation}
Unlike the $\chi\chi$ component, the supertranslation contribution here does not vanish identically. This inhomogeneous transformation dictates the required parity assignments for the reduced fields on the 2-sphere, as discussed in Section \ref{Section5}.

For the components tangent to the asymptotic 2-sphere ($\bar{A}\bar{B}$), the transformation rule for the metric perturbation ${}^1h_{\bar{A}\bar{B}}$ is directly obtained from the $\mathcal{O}(\rho)$ terms of the Lie derivative expansion \eqref{eq:Lie_gAB_expansion_app}
\begin{equation}
    \delta_\xi {}^1h_{\bar{A} \bar{B}} = \nabla_{\bar{A}} S_{\bar{B}} + \nabla_{\bar{B}} S_{\bar{A}} + 2S h_{\bar{A}\bar{B}} + \mathcal{L}_X {}^1 h_{\bar{A}\bar{B}} 
    \label{eq:delta_h_AB_app}
\end{equation}
This relation explicitly demonstrates how the supertranslation parameters $S$ and $S_{\bar{A}}$ act inhomogeneously on the angular metric perturbations, while the Lorentz sector, generated by $X$, acts via standard Lie dragging. This inhomogeneous structure is a crucial feature of the BMS group, as it describes how supertranslations ``deform'' the asymptotic geometry of the 2-sphere, ultimately determining the parity requirements for the conserved charges discussed in Section \ref{Section5}.

\end{document}